\documentclass[12pt]{article}

\usepackage{lmodern}
\usepackage[T2A,T1]{fontenc}
\usepackage[utf8]{inputenc}
\usepackage{geometry}
\usepackage{color}
\usepackage{bm}
\usepackage{mathtools}
\usepackage{amsmath}
\usepackage{amssymb}
\usepackage{graphicx}
\usepackage{subcaption}
\usepackage{textcomp}
\usepackage[numbers,sort&compress]{natbib}
\usepackage{hyperref}
\hypersetup{colorlinks=true,citecolor=black,linkcolor=black,urlcolor=black}
\makeatletter

\numberwithin{equation}{section}

\def\sect#1{Sec.~{\ref{#1}}}

\def\fig#1{Fig.~\ref{#1}}

\newcommand{\cO}{{\mathcal O}}
\newcommand{\cC}{{\mathcal C}}
\newcommand{\cM}{{\mathcal M}}
\newcommand{\ve}{{\varepsilon}}

\newcommand{\be}{\begin{equation}}
        \newcommand{\ee}{\end{equation}}
\def\nn{\nonumber}
\def\spa#1.#2{\left\langle#1\,#2\right\rangle}
\def\spb#1.#2{\left[#1\,#2\right]}
\def\sand#1.#2.#3{%
\left\langle#1{\vphantom1}\right|{#2}\left|#3\right]}%
\def\sandmp#1.#2.#3{%
\left\langle#1{\vphantom1}\right|{#2}\left|#3\right]}%
\def\sandpm#1.#2.#3{%
\left[#1{\vphantom1}\right|{#2}\left|#3\right\rangle}%
\def\sandmm#1.#2.#3{%
\left\langle#1{\vphantom1}\right|{#2}\left|#3\right\rangle}%
\def\sandpp#1.#2.#3{%
\left[#1{\vphantom1}\right|{#2}\left|#3\right]}%
\def\sandmx#1.#2.#3{%
\left\langle#1{\vphantom1}\right|{#2}\left|#3\right]}%
\def\eps{\epsilon}

\def\Tr{{\rm Tr}}

\def\tree{{\rm tree}}

\newcommand{\eqn}[1]{Eq.~\eqref{#1}}

\def\id{\protect{{1 \kern-.28em {\rm l}}}}

\newbox\charbox
\newbox\slabox
\def\s#1{{      
        \setbox\charbox=\hbox{$#1$}
        \setbox\slabox=\hbox{$/$}
        \dimen\charbox=\ht\slabox
        \advance\dimen\charbox by -\dp\slabox
        \advance\dimen\charbox by -\ht\charbox
        \advance\dimen\charbox by \dp\charbox
        \divide\dimen\charbox by 2
        \raise-\dimen\charbox\hbox to \wd\charbox{\hss/\hss}
        \llap{$#1$} }}

\allowdisplaybreaks
\begin{document}
	\interfootnotelinepenalty=10000
	\baselineskip=18pt
	\hfill
	
	\thispagestyle{empty}

	CALT-TH/2020-041
	\vspace{1.2cm}

	\begin{center}
		{ \bf \Large Leading Nonlinear Tidal Effects\\
                   and Scattering Amplitudes}

		\bigskip\vspace{1.cm}{
			{\large 
		Zvi Bern,${}^{a}$,  Julio Parra-Martinez${}^{b}$, Radu Roiban,${}^c$,  
                 \\
                Eric Sawyer${}^{a}$ and Chia-Hsien Shen,${}^{a,d}$ }
		} \\[7mm]
		{\it  
			${}^a$Mani L. Bhaumik Institute for Theoretical Physics, \\[-1mm]
			  Department of Physics and Astronomy, UCLA, Los Angeles, CA 90095, USA \\ [1mm]
			${}^b$Walter Burke Institute for Theoretical Physics,  \\[-1mm] 
                              California Institute of Technology, Pasadena, USA\\ [1mm]
			${}^c$Institute for Gravitation and the Cosmos, \\[-1mm]
			  Pennsylvania State University, University Park, PA 16802, USA \\ [1mm]
                      ${}^d$Department of Physics 0319, University of California at San Diego, \\[-1mm]
                      9500 Gilman Drive, La Jolla, CA 92093, USA
		}
                  \\
	\end{center}
	\bigskip
	\bigskip

\begin{abstract} \small

We present the two-body Hamiltonian and associated eikonal phase,  to leading post-Minkowskian 
order, for infinitely many tidal deformations described by operators with arbitrary powers 
of the curvature tensor.
Scattering amplitudes in momentum and position space provide
systematic complementary approaches.
For the tidal operators quadratic in curvature, which describe the linear response to
an external gravitational field, we work out the leading post-Minkowskian contributions 
using a basis of operators with arbitrary numbers of derivatives 
which are in one-to-one  correspondence
with the worldline multipole operators. 
Explicit examples are used to show that the same techniques apply to
both bodies interacting tidally with a spinning particle, for which we find the leading 
contributions from quadratic in curvature tidal operators with an arbitrary number 
of derivatives, and to effective field theory extensions of general relativity. 
We also note that the leading post-Minkowskian order contributions from higher-dimension 
operators manifest double-copy relations.
Finally, we comment on the structure of higher-order corrections.

\end{abstract}


	\setcounter{footnote}{0}
	
\renewcommand{\baselinestretch}{1}	
	\newpage
	\setcounter{tocdepth}{2}
	\tableofcontents
	
	\newpage

\section{Introduction}
\label{IntroSection}

The remarkable discovery of gravitational waves by the LIGO and
Virgo collaborations~\cite{LIGO} has ushered in a new era of
exploration that promises major new discoveries on black holes,
neutron stars and perhaps even new basic insights into fundamental physics.
Theoretical tools of increased precision, matching that of gravitational-wave 
signals not only from current detectors but also from proposed gravitational-wave 
observatories~\cite{NewDetectors}, are required.

The evolution of a compact binary and the ensuing gravitational-wave
emission can be divided in three distinct phases --- inspiral, merger and ring down --- according 
to their underlying properties. 
The inspiral part of binary mergers, which is the subject of this paper, is analyzed through 
models such as the effective one-body (EOB) formalism \cite{EOB}.  The weak gravitational 
field during this phase makes it suitable for a perturbative approach and these models import 
information from post-Newtonian~(PN) gravity~\cite{PN, NRGR, PN_EFT, 6PNCrossCheck}, as well as
the self-force framework~\cite{selfforce} and numerical relativity~\cite{NR}.
More recently, the post-Minkowskian~(PM) expansion~\cite{PM, CliffIraMikhailClassical, 3PMPRL, 
3PMLong,  Antonelli:2019ytb, CheungSolon3PM, Porto3PM} has gained prominence due to 
its capture of the complete velocity dependence at fixed order in Newton's constant.  
By exposing the analytic structure of each order, this expansion also offers new insight into features of 
gravitational perturbation theory, exposes hereto unexpected structure in certain observables, and 
may open a path to the resummation of perturbation theory in the classical limit.
The PN, PM and self-force expansions provide important nontrivial cross checks in their overlapping 
regions of validity~\cite{3PMLong, Antonelli:2019ytb, 6PNCrossCheck, SelfForceComparisions}.  
For recent reviews see Refs.~\cite{GravityReviews}.

Over the years a close link between classical physics and scattering amplitudes has been 
developed~\cite{ScatteringToClassical, DamourTwoLoop, CliffIraMikhailClassical, 3PMPRL, 
3PMLong, CheungSolon3PM, CheungSolonTidal} and 
led to a robust  and powerful means for obtaining two-body Hamiltonians~\cite{CliffIraMikhailClassical}
and observables in the post-Minkowskian expansion. It was obtained by combining
modern techniques, such as generalized unitarity \cite{GeneralizedUnitarity},  which emphasize 
gauge-invariant building blocks at all stages and build higher-order contributions from lower-order  
ones with effective field theory methods.
This framework proved its effectiveness 
through the construction of the 
sought after two-body Hamiltonian at the third order in Newton's constant~\cite{3PMPRL, 3PMLong} and the 
identification of surprising simplicity in physical observables of interacting spinning black 
holes~\cite{Bern:2020buy}. The scattering angle is of particular importance,
as it provides a direct link~\cite{DamourTwoLoop} with the EOB framework~\cite{EOB} used 
to predict gravitational wave emission from compact binaries.

In this paper we investigate the effects of tidal deformations~\cite{TidalReview} on 
the conservative two-body Hamiltonian during the inspiral phase, focusing on their 
structure in the post-Minkowskian expansion.
The tidal deformations offer a window into the equation of state of neutrons
stars \cite{NeutronStars} and test our understanding of black holes~\cite{DamourTidal, 
PNTidal, PMTidal, CheungSolonTidal, PortoTidalTwoLoop, 
Haddad:2020que} and of possible exotic physics~\cite{UnusualPhysicsTidal}.  
While tidal effects are expected to vanish for black holes in general
relativity~\cite{LoveZero}, they are of crucial importance for understanding the equation 
of state of neutron stars.  These corrections are formally equivalent to fifth-order
post-Newtonian effects~\cite{NRGR}, highlighting the importance of precision 
perturbative calculations.

Properties of extended bodies that relate to their finite size can be encoded in 
local-operator deformations of a point-particle theory by integrating out their internal 
degrees of freedom. The set of all possible tidal operators is constrained only by the 
symmetry properties of the fundamental theory, such as parity.
We introduce our organization of tidal operators in close analogy with the case 
of electromagnetic susceptibilities. Indeed, not only is there a formal similarity between 
gauge theory and gravity, but 
the integrand of gravitational scattering amplitudes can be obtained directly from 
gauge theory using the double copy~\cite{KLT,BCJ}.  
For the relatively simple case of the leading-PM order contribution
of a given tidal operator to scattering amplitudes,  these relations follow from 
the factorization of the point-particle energy-momentum tensor and 
from the fact that the linearized Riemann tensor is a product of two 
gauge-theory field strengths.
Thus, in analogy with the case of electromagnetic interactions of extended bodies, 
tidal operators may contain arbitrarily-high number of Riemann curvature tensors 
with an arbitrary number of derivatives.

Curvature-squared tidal operators,
 describing the linear response of an extended body to an external 
gravitational field, were recently classified in Ref.~\cite{Haddad:2020que}, where an expression
for the two-body Hamiltonian and scattering angle at leading post-Minkowskian
order was conjectured.
Here we prove the conjecture for a basis of operators whose Wilson coefficients
in the four-dimensional point-particle effective action are exactly the same as the worldline electric 
and magnetic tidal coefficients, related to the corresponding multipole Love numbers 
by factors of the typical scale of the body, see e.g. Ref.~\cite{NRGR, PMTidal, NeutronStars}. 
The lowest-order matrix elements of our tidal operators are, by construction,
the same as the matrix elements of the worldline tidal operators. 
To establish the map beyond leading order it is necessary to compare physical 
quantities. At the next-to-leading order the contributions of low-derivative $R^2$ tidal 
operators to the two-body Hamiltonian and to the scattering angle were determined in
Refs.~\cite{CheungSolonTidal,PortoTidalTwoLoop}.

We also obtain the leading-order modifications of the two-body Hamiltonian and 
of the scattering angle due to tidal operators with arbitrarily-high number of Weyl tensors, 
which describe the nonlinear response of extended bodies to external gravitational field. 
As usual we organize the operators in terms of electric and magnetic-type components, $E$ and $B$,
of the Riemann (or Weyl) tensor.  The finite rank of these tensors leads to nontrivial relations 
between different operators, allowing us to express the contributions of $E^n$ and $B^n$-type operators 
for $n\ge 4$ in terms of those of products of simpler operators, thus reducing the number of independent 
structures.

While these relations appear mysterious for scattering amplitudes in momentum space,
they are made manifest by Fourier-transforming the integral representation of the 
amplitude to position space. At any loop order, the transform decouples all integrals 
from each other.  This observation allows us to write down closed-form expressions for 
amplitudes, two-body Hamiltonians and scattering angles generated by 
infinite families of operators.
Beyond leading order the structure of tidally-deformed amplitudes is more complicated, 
but the momentum-space methods of Refs.~\cite{CliffIraMikhailClassical, 3PMPRL,
3PMLong,CheungSolonTidal} can be applied systematically. Integration by parts 
methods~\cite{IBP} are especially powerful for the conservative two-body problem 
because in the potential region of loop integrals all relevant integrals are of 
single-scale type~\cite{Parra-Martinez:2020dzs}.

The methods we use to describe tidal operators apply equally well to deformations
of a point-particle theory by any operators, including e.g. those arising in effective
field theory extensions of General Relativity~\cite{GravityEFT,pureR4,pureR3,pureR3Other}. 
We illustrate this point by working out the contributions of $R^3$ and $R^4$ and 
compare them with existing results.  
The two-body Hamiltonian and associated observables for a point-particle deformed 
by tidal operators interacting with a spinning particle can also be derived through similar 
methods. To leading PM order, only the single-graviton interaction of the spinning 
particle is relevant and it is captured by the stress tensors described in~\cite{LeviSteinhoffLagrangian, 
Vines:2017hyw, Bern:2020buy}. As an example, we find the leading spin-orbit 
contributions from $E^2$-type tidal operators with an arbitrary number of derivatives 
interacting with a spinning particle.

This paper organized as follows. 
In~\sect{OperatorsSection} we present a description of the operators encoding tidal deformations.  
In~\sect{LOR2Section} we discuss the leading-order tidal contributions from $R^2$-type operators 
with an arbitrary number of derivatives.  This section also demonstrate how to incorporate spin effects 
for the second body.  
We proceed to derive  in \sect{LORnSection} the leading contributions of various infinite classes of $R^n$-type tidal 
operators and also comment on their higher-order contributions.  In~\sect{LOPureRnSection} we discuss the application
of our methods to the case of $R^n$ extensions of General Relativity.  We present our conclusions in \sect{ConclusionSection}.  
An appendix gives the explicit results for the contributions of a collection of high-order tidal operators 
to the two-body Hamiltonian and the associate scattering amplitudes.

\vskip .3 cm {\bf Note added:} While this project was ongoing we
became aware of concurrent work by Cheung, Shah and
Solon~\cite{CliffordMikhailNewTidal} based on using the geodesic
equation and containing some overlap on leading contributions to the
two-body Hamiltonian from the $R^n$ tidal operators.  In addition, the
methods developed there determine the two-body Hamiltonian for a
tidally-deformed test particle interacting with a Schwarzschild black
hole, to all orders in the Schwarzschild radius of the latter.
We are grateful for interesting and helpful discussions and sharing drafts.

\section{Effective actions for tidal effects}
\label{OperatorsSection}

\subsection{Effective actions for post-Minkowskian potentials}

In this work we study tidal or finite-size effects in the gravitational interactions of two massive extended bodies. 
They are encoded in a classical two-body Hamiltonian of the form
\begin{equation}
\label{hamiltonian}
  H(\bm p , \bm r) = \sqrt{\bm p^2 + m_1^2} + \sqrt{\bm p^2 + m_2^2} + V(\bm p, \bm r)\, ,
\end{equation}
and is extracted systematically, following the general approach introduced in~\cite{CliffIraMikhailClassical}, by matching QFT 
scattering amplitudes to a non-relativistic EFT.  
If the size of the two bodies is much smaller than their separation, 
non-analytic/long-distance classical potential has the form
\begin{equation}
  V(\bm p, \bm r) \sim c_{i}(\bm p) \, m \left(\frac{Gm}{|\bm r|}\right)^{i}  \, ,
\end{equation}
where $m$ carries unit mass dimension and 
the momentum transfer $\bm q$, Fourier-conjugate to $\bm r$, is much smaller than the center of mass momentum $\bm p$.
Such a conservative potential arises from integrating out gravitons with momenta $\ell$ in the potential region which has the scaling behavior
\begin{equation}
\ell = (\ell^0, \bm \ell) \sim (|\bm q| |\bm v|, |\bm q|),
\end{equation}
where $|\bm v|\sim \mathcal{O}\left(|\bm p|/m \right)$.  Note that
$Gm$ is of the order of the effective Schwarzschild radius of the
particles $R_s$, so the classical expansion\footnote{The amplitude
  also contains non-analytic terms which we will not study here,
  corresponding to quantum contributions to the potential of the form
  $(\ell_p^2/r^2)^n$, where $\ell_p$ is the Planck length.} of the
potential is an expansion in $R_s/|\bm r|$. If the separation of the
two bodies can be of the same order as their typical size $R$, then
the classical potential takes the form
\begin{equation}
  V(\bm p, \bm r) \sim c_{i,k}(\bm p) \, m \left(\frac{Gm}{|\bm r|}\right)^{i} \left(\frac{R}{|\bm r|}\right)^k \, .
\end{equation}
For black holes $R\sim R_s$ so the size of terms with powers of $R$ is
comparable to higher PM orders. For other bodies $R>R_s$ so the
contribution should be bigger. For reference, neutrons stars have
$R/R_s \sim 10$, and the sun has $R/R_s \sim 10^5$. In practice, it is
convenient to always use $R_s/r$ as the expansion parameter so that
the tidal effects just modify the coefficients in the usual PM
potential, i.e. $c_{i,k} \sim \Delta c_{i+k}$.

From our point of view, the new scale $R_s$ is introduced by
integrating out the degrees of freedom that describe the tidal
dynamics of an extended body to yield a point-particle effective
theory. In such an effective theory the finite size effects are
encoded as higher-dimension operators ${\cal O}_i$ which are
suppressed by powers of $R_s |\bm q|$. Their Wilson coefficients can
be determined either by matching to the complete theory that includes
the tidal degrees of freedom, or by comparing to experiment. A side
effect of choosing $R_s$ instead of $R$ as the scale characterizing
finite-size effects is that for less compact bodies the Wilson
coefficients are not necessarily $\mathcal{O}(1)$.  This approach was
pioneered in the context of a worldline PN formalism in
Ref.~\cite{NRGR}, and recently adapted to the PM framework in
Ref.~\cite{PortoTidalTwoLoop}. In the QFT language this approach has
been recently used in Refs.~\cite{CheungSolonTidal, Haddad:2020que}.
In section we provide a systematic treatment of such effective actions
and write a basis of operators which simplifies the translation
between QFT and worldline formalisms and makes the relation to
familiar \emph{in-in} observables manifest.

The cases that we focus on in this paper correspond to leading
contributions from tidal or other operators. Although these operators
first contribute to loop amplitudes, the determination of their
leading-order contribution to the two-body potential is
straightforward and formally given by inverting the Born relation
between the scattering amplitude and the potential:
\begin{equation}
\label{VvsM}
  V_{\mathcal{O}}(\bm p,\bm r) = -\frac{1}{4 E_1 E_2} \int {d^{D-1} \bm q \over (2 \pi)^{D-1}} e^{-i \bm {q} \cdot \bm r}   \mathcal{M}_{\mathcal{O}}(\bm p,\bm q) \, .
\end{equation}
Here $\mathcal{M}_{\mathcal{O}}$ is the leading-order four-scalar scattering amplitude with with a single insertion of $\mathcal{O}$, 
center of mass momentum $\bm p$, transferred momentum $\bm q$.
In general the potential is gauge dependent and not unique. In the above equation we choose to expose the on-shell condition on $\bm q$ first such that $\bm p\cdot \bm q \simeq\mathcal{O}(\bm q^2)\sim 0$. This naturally gives the potential in the isotropic gauge.

Alternatively, the effective two-body Hamiltonian can be constructed
by matching its conservative observables --- such as the conservative
scattering angle, or the impulse and spin kick --- or the
closely-related eikonal phase~\cite{EikonalPapers},
\begin{align}
  \delta_{\mathcal{O}}(\bm p, \bm b) &= \frac{1}{4m_1 m_2 \sqrt{\sigma^2-1}} \int  \frac{d^{D-2}\bm q}{(2\pi)^{D-2}} e^{-i\bm b\cdot \bm q} \mathcal{M}_{\mathcal{O}}(\bm p,\bm q) \,,
  \label{PhasevsM}
\end{align}
with the corresponding quantities  in the complete theory.
Here we use $-p_i=-m_i u_i$ as the incoming momenta of particle 1 and 2 and
\begin{align}
\sigma \equiv \frac{p_1\cdot p_2}{m_1 m_2} = u_1\cdot u_2\,.
\end{align}
In either case, the matching is carried out order by order in Newton's
constant $G$, that is order by order in the post-Minkowskian
expansion. The relation between the eikonal and conservative
observables holds also for the scattering of spinning particles. To
leading nontrivial order, the effect of a composite operator ${\cal
  O}$ on the impulse and spin kick in the center-of-mass frame is
\begin{align}
\Delta \bm p = -\nabla_{\bm b} \delta_{\mathcal{O}}(\bm b) +\dots \,,
\qquad
\Delta \bm S_i = -\{{\bm S}_i, \delta_{\mathcal{O}}(\bm b)\} +\dots \,,
\label{spinningeikonal}
\end{align} 
where the ellipsis stand for higher-order terms that depend on ${\cal
  O}$ and $\{\bullet,\bullet\}$ is the Poisson bracket. We expect that
the all-order relation between the eikonal phase and conservative
observables put forth in Ref.~\cite{Bern:2020buy} holds in the
presence of deformations by tidal and other composite operators.
At leading order, the semiclassical approximation implies that the eikonal
phase coincides with the radial action integrated over the scattering
trajectory.  We discuss this further in Section~\ref{real_oneloop}.  The
latter allows us to make contact with Ref.~\cite{PMTidal} in which tidal
effects were computed using a classical worldline formalism for a subset of
tidal operators.

Alternatively the matching can be performed by directly computing a
physically meaningful quantity such as the conservative scattering
angle, corresponding to the scattering with radiation reaction turned
off; or the closely related eikonal phase.  In either case matching is
performed order by order in perturbation theory in Newton's constant,
$G$, that is order by order in the post-Minkowskian expansion.
\begin{align}
\label{eq:Mbardef}
\mathcal{M}_{\mathcal{O}}(\bm q) &=  |\bm q|^{A} \overline{\mathcal{M}}_{\mathcal{O}} \,, \\
\label{eq:VMbar}
V_{\mathcal{O}}(\bm r) &= -\frac{1}{4E_1E_2} \frac{2^A \Gamma\left(\frac{1}{2}(D-1 + A)\right)}
   {  \pi^{(D-1)/2}\Gamma(-\frac{1}{2}A)} |\bm r|^{-A-(D-1)} \, \overline{\mathcal{M}}_{\mathcal{O}} \,,\\
\delta_{\mathcal{O}}(\bm b) &= \frac{1}{4m_1 m_2 \sqrt{\sigma^2-1}} 
\frac{2^A \Gamma\left(\frac{1}{2}(D-2 + A)\right)}{  \pi^{(D-2)/2}\Gamma(-\frac{1}{2}A)} |\bm b|^{-A-(D-2)} \,
\overline{\mathcal{M}}_{\mathcal{O}} \, , 
\label{phaseV1}
\end{align}
where we have used the formula for the Fourier transform of a power
\begin{equation}
  \int \frac{d^D \bm q }{(2\pi)^D}  e^{-i \bm x \cdot \bm q} |\bm q|^A = 
\frac{2^A  \Gamma\left(\frac{1}{2}(D + A)\right)}{ \pi^{d/2}\Gamma(-\frac{1}{2}A)} 
  |\bm x|^{-(A+D)} \,.
  \label{eq:FTransform}
\end{equation}
Here $A$ is power of the soft $q$ carried by the amplitude. For an
operator with $n$ power of Riemann or Weyl tensors with $n_{\partial}$
derivatives acting on them, the leading contribution to the two-to-two
scalar amplitude is
\begin{equation}
  A = 3n+n_{\partial}-3-2\eps(n-1),
\end{equation}
where we use $D=4-2\eps$.  For example, for the electric and magnetic
operators $E^2$ and $B^2$ we will introduce shortly, $n=2$ and
$n_{\partial}=0$ so $A=3-2\eps$, and every pair of derivatives acting
of these increases $n_{\partial}$ and $A$ by two.

\subsection{Effective actions for linear and non-linear tidal effects}

We now explain how to parametrize the response of a general body to
an external field and how this can be encoded in an effective action.
We will discuss this in detail in the simpler case of
electromagnetism, which will easily generalize to the gravitational
case.

\subsubsection{Tidal response in non-linear optics}

The full non-linear response of a body to an external electric field $E_i$ is described by the induced electric dipole moment density $D_i$. In the rest frame of the body, it has a formal expansion in powers of the electric field~\cite{Boyd}:
\begin{align}
  D_{i_1}(t,\bm x) &= \chi^{(1)}_{i_1i_2}(t,\bm x)E_{i_2}(t,\bm x) + \chi^{(2)}_{i_1i_2i_3}(t,\bm x)E_{i_2}(t,\bm x)E_{i_3}(t,\bm x)+ \cdots \, .
  \label{eq:Dx}
\end{align}
The first term is the familiar linear response function; the
subsequent terms encode the properties of the body in the
susceptibility tensors, $\chi^{(n)}$, which are symmetric in their
indices. Similarly, in the presence of a magnetic field $B_i$, one can
write magnetic susceptibilities, as well as general susceptibilities
capturing the response under a general electromagnetic field.

It is convenient to transform Eq.~\eqref{eq:Dx} to Fourier space, where it takes the form
\begin{align}
  D_{i_1}(-\omega_1,-\bm q_1) &= \chi^{(1)}_{i_1i_2}(\omega_1,\bm q_1;\omega_2,\bm q_2)E_{i_2}(\omega_2,\bm q_2) \nonumber\\
  & \,\hskip .3 cm \null
   + \chi^{(2)}_{i_1i_2i_3}(\omega_1,\bm q_1;\omega_2,\bm q_2;\omega_3,\bm q_3)E_{i_2}(\omega_2,\bm q_2)E_{i_3}(\omega_3,\bm q_3)  + \cdots\, .
  \label{eq:Dq}
\end{align}
Here we have adopted a generalized summation convention where repeated
frequencies and momenta are integrated over, and the Fourier
susceptibilities include energy-momentum-conservation delta functions
\begin{equation}
  \chi^{(n-1)}_{i_1\cdots i_n} = \delta\left(\sum_i \omega_i\right) \delta\left(\sum_i \bm q_i\right)
  \tilde\chi^{(n-1)}_{i_1\cdots i_n}\,,
  \label{eq:momcons1}
\end{equation}
which account for the fact that the position-space product in Eq.~\eqref{eq:Dx} becomes a Fourier space convolution in Eq.~\eqref{eq:Dq}.

The dipole density can be related to a generating function --- or effective action --- $S(E)$, via the usual response formula
\begin{equation}
D_{i_1}(-\omega_1,-\bm q_1) = \frac{\partial S(E)}{\partial E^{i_1}(\omega_1,\bm q_1)} \, .
\end{equation}
The effective action, following from formally integrating Eq.~\eqref{eq:Dx}, is given by
\begin{align}
  S(E) &=  \frac12 \chi^{(1)}_{i_1i_2}(\omega_1,\bm q_1;\omega_2,\bm q_2)  E_{i_1}(\omega_1,\bm q_1)E_{i_2}(\omega_2,\bm q_2)  \nonumber \\
  &+ \frac13 \chi^{(2)}_{i_1i_2i_3}(\omega_1,\bm q_1;\omega_2,\bm q_2;\omega_3,\bm q_3)  E_{i_1}(\omega_1,\bm q_1) E_{i_2}(\omega_2,\bm q_2)E_{i_3}(\omega_3,\bm q_3) + \cdots\,.
\end{align}
This makes clear that the momentum space susceptibilities are completely symmetric tensors, as well as symmetric functions of all their arguments. 
$S(E)$ could be put in a form closer to an action by series expanding the susceptibilities and rewriting the powers of frequency and three-momenta as derivatives. For instance one can rewrite some terms in the expansion as follows
\begin{align}
  &\left(\frac{\partial \chi^{(1)}_{i_1i_2}}{\partial\omega_1\partial\bm q_2^j}(0) \, \omega_1 \bm q_2^j \right)  E_{i_1}(\omega_1,\bm q_1)E_{i_2}(\omega_2,\bm q_2) \sim  \left(\frac{\partial \chi^{(1)}_{i_1i_2}}{\partial\omega_1\partial\bm q_2^j}(0) \right) \partial_t E_{i_1}(t,\bm x) \bm\nabla^j_{\bm x}E_{i_2}(t,\bm x)\,.
\end{align}
Note that the expansion in the three momenta here simply corresponds to a multipole expansion of the electric fields. 

So far we have been working in the rest frame of the object. The choice of a frame breaks manifest Lorentz invariance 
down to the rotations around the position of the object.
We would like to covariantize the expressions above so that is they are valid in an arbitrary reference frame, in which the 
body moves with velocity $\bm v$. This can be done by considering the four-velocity of the object $u^\mu = \gamma (1,\bm v)$, 
where $\gamma$ is the Lorentz factor. As is well known the electric field and magnetic fields in the rest frame of the body 
can be covariantly written as
\begin{equation}
  E_\mu = F_{\mu\nu}u^\nu \,,  \quad B_\mu = *F_{\mu\nu}u^\nu \,, 
\end{equation}
where $F_{\mu\nu}$ is the electromagnetic field strength, and $*F_{\mu\nu}$ its dual. 
Similarly, it is clear that any frequency and spatial momenta can be written as
\begin{equation}
  \omega_i \rightarrow  u \cdot q  \equiv u^{\mu} q_{\mu} \,, \qquad \bm q_i \rightarrow  (q^\perp)_{\mu} \equiv P_{\mu\nu} q^\nu\,,
\end{equation}
where we have introduced the four momentum of the field, $q_i^\mu$ and a projector,
\begin{equation}
  P^{\mu\nu} = \eta^{\mu\nu} - u^{\mu}u^{\nu} \,,
\end{equation}
which makes indices purely spatial in the rest frame of the object.
Naively this covariantization requires adding components to the polarizabilities so that $\chi^{(n-1)}_{i_1\cdots i_n}\rightarrow \chi^{(n-1)}_{\mu_1\cdots \mu_n}$, and we can write
\begin{align}
S(E) &= \chi^{(1)}_{\mu_1\mu_2}(u\cdot q_1,  q^\perp_1;u\cdot q_2, q^\perp_2)  E_{\mu_1}(q_1)E_{\mu_2}(q_2)  \nonumber \\
&+ \chi^{(2)}_{\mu_1\mu_2\mu_3}(u\cdot q_1, q^\perp_1;u\cdot q_2, q^\perp_2,u\cdot q_3,  q^\perp_3) E_{\mu_1}(q_1)E_{\mu_2}(q_2)E_{\mu_3}(q_3)  +\cdots \,,
\end{align}
due to the fact that $u^{\mu}E_{\mu} = u^{\mu}B_{\mu} = 0$, which follows from the antisymmetry of the field strength.

The generating function written above describes the non-linear
response of an arbitrary material, including those that violate
rotational and Lorentz invariance. In the following we will be only
interested in Lorentz-preserving effects, which impose addition
constraints on the susceptibility tensors. Firstly, Lorentz
invariance constrains the index structure of the susceptibility, which
can only be carried by Lorentz-covariant tensors. If we impose parity,
the only such tensors are the metric itself and the graviton momenta,
so the tensor susceptibility must decompose in a set of scalar
susceptibilities as follows
\begin{align}
\label{LIconstraints1}
\chi^{(1)}_{\mu_1\mu_2}&= \chi^{(1)}_0 \, g_{\mu_1\mu_2}   + \chi^{(1)}_1 \, q^\perp_{1\,\mu_1} q^\perp_{2\,\mu_2} \\
  \chi^{(2)}_{\mu_1\mu_2\mu_3}&= \chi^{(2)}_0 \,( g_{\mu_1\mu_2}q^\perp_{3\,\mu_3}   + g_{\mu_2\mu_3}q^\perp_{1\,\mu_1} + g_{\mu_3\mu_1}q^\perp_{2\,\mu_2}) \,,
  \\
   \chi^{(3)}_{\mu_1\mu_2\mu_3\mu_4}&= \chi^{(3)}_0 \, g_{(\mu_1\mu_2}  g_{\mu_3\mu_4)} 
   + \chi^{(3)}_1 \, (g_{\mu_1\mu_2}  q^\perp_{3\mu_3} q^\perp_{4\mu_4} + \text{perms}) 
   + \chi^{(3)}_2 \, q^\perp_{1\mu_1} q^\perp_{2\mu_2} q^\perp_{3\mu_3} q^\perp_{4\mu_4} \,,
\label{LIconstraintsn}
\end{align}
where in general each tensor structure must be summed over
permutations which respect the symmetry $(\mu_i\leftrightarrow \mu_j)$
while simultaneously swapping $q^\perp_i \leftrightarrow q^\perp_j$.
Another consequence of Lorentz invariance is that the scalar
susceptibilities only depend on Lorentz invariant combinations of
momenta, so that
\begin{equation}
  \chi^{(n-1)}_{a}(u\cdot q_i;  q^\perp_i) \rightarrow \chi^{(n-1)}_{a}(u\cdot q_i ;  q_i^\perp\cdot q_j^\perp)\,.
\end{equation}
Note that in the rest frame $q_i^\perp\cdot q_j^\perp = \bm q_i\cdot \bm q_j$.

\subsubsection{Non-linear tidal response in gravity}

It is now easy to generalize the tidal response for electromagnetism
to its gravitational analog. In this case we start from the induced
quadrupole moment, written in terms of the gravito-electric field
\begin{align}
  Q_{i_1j_1}(t,\bm x) &= \chi^{(1)}_{i_1j_1i_2j_2}(t,\bm x)E_{i_2j_2}(t,\bm x) + \chi^{(2)}_{i_1 j_1 i_2j_2i_3j_3}(t,\bm x)E_{i_2j_2}(t,\bm x)E_{i_3j_3}(t,\bm x)+ \cdots \,,
  \label{eq:Qx}
\end{align}
where now the gravitational susceptibilities are more general tensors symmetric in each pair of $i$ and $j$ indices
\begin{align}
 \chi_{\cdots i j \cdots} = \chi_{\cdots ji \cdots}\,,  \hskip 1.8 cm 
 \chi_{\cdots i_a j_a \cdots i_b j_b  \cdots} = \chi_{\cdots i_b j_b \cdots i_a j_a  \cdots}\, .
\end{align}
In the rest frame of the object the electric field is related to the Weyl tensor as $E_{ij} = C_{0i0j}$. Similar expressions 
can be written for the response to a gravito-magnetic or to a mixed field.

All of these quantities can be covariantized by introducing
\begin{eqnarray}
E_{\mu\nu}\equiv C_{\mu\alpha\nu\beta} u^{\alpha}u^{\beta},
\qquad
B_{\mu\nu}\equiv (*C)_{\mu\alpha\nu\beta} u^{\gamma}u^{\delta}\equiv \frac{1}{2}\epsilon_{\alpha\beta\gamma\mu} C^{\alpha\beta}_
{\quad\delta\nu} u^{\gamma}u^{\delta},
\label{eq:EandB_Def}
\end{eqnarray}
where all indices are curved and the Levi-Civita tensor is defined as $\epsilon^{0123} = +1$. As in the electromagnetic case the following relations hold
\begin{equation}
  E_{\mu\nu} u^{\nu}  =0 \,, \qquad B_{\mu\nu} u^{\nu} = 0\,,
  \label{eq:EBnull}
\end{equation}
as well as
\begin{equation}
  E_{\mu}{}^{\mu} = 0 \,, \qquad B_{\mu}{}^{\mu} =0\,,
\end{equation}
where the first equality is a consequence of the tracelessness of the Weyl tensor.
The corresponding generating function for tidal response is then simply 
\begin{align}
  S_{\text{grav} }(E) &=  \chi^{(1)}_{\mu_1\nu_1\mu_2\nu_2}(u\cdot q_1, Pq_1;u\cdot q_2,Pq_2)  \phi(p')E^{\mu_1\nu_1}(q_1)E^{\mu_2\nu_2}(q_2)\phi(p)   \\
& \hskip .3 cm \null
+ \chi^{(2)}_{\mu_1\nu_2\mu_2\nu_2\mu_3\nu_3}(u\cdot q_1, Pq_1;u\cdot q_2,Pq_2,u\cdot q_3, Pq_3) \phi(p')E^{\mu_1\nu_1}(q_1)E^{\mu_2\nu_2}(q_2)E^{\mu_3\nu_3}(q_3) \phi(p) + \cdots \nonumber
\end{align}
where, as above, a convolution over all momenta is implicit, and the covariant susceptibilities are traceless in each pair of $\mu,\nu$ indices $\eta^{\mu\nu}\chi_{\cdots\mu\nu\cdots}=0$. Once again, Lorentz invariance will further constraint the form of the susceptibility tensors in a way analogous to Eqs.~\eqref{LIconstraints1}-\eqref{LIconstraintsn}.

\subsubsection{From response to QFT effective actions}

We now proceed to connect our discussion to a QFT effective action, focusing on the case of gravity; the electromagnetic case is 
completely analogous.

The connection can be easily made by interpreting the generating
function, $S_{\text{grav} }(E)$ as the expectation value in a
background field of an operator in a one-particle state $| p \rangle$
with four momentum $p = m u$, and zero spin.
In second-quantized language the one-particle state is created by a scalar field, $\phi$, at infinity and 
\begin{align}
  S_{\text{tidal}} &=    \chi^{(1)}_{\mu_1\mu_2}(u\cdot q_1,  q^\perp_1;u\cdot q_2, q^\perp_2)   \phi(p) E_{\mu_1}(q_1)E_{\mu_2}(q_2) \phi(p') \nonumber \\
&+ \chi^{(2)}_{\mu_1\mu_2\mu_3}(u\cdot q_1,  q^\perp_1;u\cdot q_2, q^\perp_2,u\cdot q_3,  q^\perp_3)   \phi(p) E_{\mu_1}(q_1)E_{\mu_2}(q_2)E_{\mu_3}(q_3) \phi(p') +\cdots \,,
\end{align}
can be identified as the momentum-space effective action that encodes the response to the background field. Note that, in order to enforce momentum conservation, the Fourier-transformed susceptibilities must satisfy 
\begin{equation}
  \chi^{(n-1)}_{\mu_1\cdots \mu_n} = \delta\left(\sum_i q_i - q\right)   \tilde\chi^{(n-1)}_{\mu_1\cdots \mu_n}\,,
  \label{eq:momcons2}
\end{equation}
where $q=-(p+p')$. Note that the susceptibilities are initially only defined for $q=0$, so their covariantization requires an extension to $q\neq 0$. This does not affect the classical limit.  As above, each term in the expansion of susceptibilities is encoded by a higher-dimension operator 
in the effective action, where now the factors of four-velocity $u$ can be identified with derivatives acting on the scalar field. 
For instance,
\begin{align}
  &\partial_{\omega}^{2n}\chi^{(1)}_{\mu_1\nu_1\mu_2\nu_2}(0,0) [(u\cdot q_1)^{2n} + (u\cdot q_2)^{2n}]  \phi(p')E^{\mu_1\nu_1}(q_1)E^{\mu_2\nu_2}(q_2)\phi(p) \nn \\,
  &\hspace{2cm} \leftrightarrow   \partial_{\omega}^{2n}\chi^{(1)}_{\mu_1\nu_1\mu_2\nu_2}(0,0) \int d^4x \sqrt{-g}\,\frac{1}{m^{2n}}  \phi E^{\mu_1\nu_1} \nabla_{(\rho_1\cdots\rho_{2n})}E^{\mu_2\nu_2}\nabla^{(\rho_1\cdots\rho_{2n})}\phi \,.
\end{align}
where the classical limit is implicit on the left-hand side.
To write a generic operator appearing in this expansion it is convenient to introduce the combinations,
\begin{align}
\label{QFToperators}
{\hat E}_{\mu_1\mu_2\dots\mu_n} &=\frac{i^2}{m^2}{\rm Sym}_{\mu_1\dots \mu_n}[\nabla_{\nu_n}\dots\nabla_{\nu_3} C_{\mu_1\alpha\mu_2\beta} {\hat P}_{\mu_n}^{\nu_n}\dots  {\hat P}_{\mu_3}^{\nu_3}\nabla^\alpha\nabla^\beta \; ] \,,
\nn \\
{\hat B}_{\mu_1\mu_2\dots\mu_n} &= \frac{i^2}{m^2}{\rm Sym}_{\mu_1\dots \mu_n}[\nabla_{\nu_n}\dots\nabla_{\nu_3} (*C)_{\mu_1\alpha\mu_2\beta} {\hat P}_{\mu_n}^{\nu_n}\dots  {\hat P}_{\mu_3}^{\nu_3}\nabla^\alpha\nabla^\beta \; ] \,,
\nn \\
{\hat { E}}{}^{(l)}_{\mu_1\mu_2\dots\mu_n} &= \frac{i^{m+2}}{m^{m+2}}
{\rm Sym}_{\mu_1\dots \mu_n}[\nabla_{\nu_n}\dots\nabla_{\nu_3} \nabla^{\rho_1}\dots  \nabla^{\rho_l} 
 C_{\mu_1\alpha\mu_2\beta} {\hat P}_{\mu_n}^{\nu_n}\dots  {\hat P}_{\mu_3}^{\nu_3}\nabla_{(\rho_1}\dots \nabla_{\rho_l)}  \nabla^\alpha\nabla^\beta \; ]\,,
\nn \\
{\hat { B}}{}^{(l)}_{\mu_1\mu_2\dots\mu_n} &= \frac{i^{m+2}}{m^{m+2}}{\rm Sym}_{\mu_1\dots \mu_n}[
\nabla_{\nu_n}\dots\nabla_{\nu_3}  \nabla^{\rho_1}\dots  \nabla^{\rho_l} (*C)_{\mu_1\alpha\mu_2\beta} {\hat P}_{\mu_n}^{\nu_n}\dots  {\hat P}_{\mu_3}^{\nu_3}\nabla_{(\rho_1}\dots \nabla_{\rho_l)} \nabla^\alpha\nabla^\beta \; ]\, ,
\end{align}
where all the derivatives on the right of the Weyl tensor act on the scalar field, and the position-space projector is
\begin{equation}
{\hat P}_{\mu}^{\nu}=\frac{1}{m^2}(\partial_\mu\partial^\nu - \delta_\mu^\nu \partial^2) \, .
\label{u_projector}
\end{equation}
The terms in the expansion that encode the most general linear response are then
\begin{align}
  S^\text{QFT}_\text{tidal}\big|_{\text{linear}}= m \int d^4x \sqrt{-g}
&\hphantom{+} \! \sum_{n=2}^\infty \sum_{l =0}^\infty (\mu^{(n,l)} \, \phi \hat E^{(l)}_{\mu_1\cdots\mu_n} \hat E^{(l)\,\mu_1\cdots\mu_n} \phi
+\sigma^{(n,l)} \, \phi \hat B^{(l)}_{\mu_1\cdots\mu_n} \hat B^{(l)\,\mu_1\cdots\mu_n}\phi)\,
\end{align}
where the coefficients are related to the susceptibility as 
$\mu^{(n,l)}\sim (\partial_{\omega_2})^l(\partial_{q_1\cdot q_2})^l \chi^{(1)}_0(0;0)$, 
and the magnetic susceptibilities are related to $\sigma^{(n,l)}$ in a similar way.
Operators like $\phi E_{\mu_1\mu_2}^{(l_1)} E^{(l_2)}{}^{\mu_1\mu_2}\phi$ with
$l_1\ne l_2$  are related to operators with $l_1=l_2$ by integration by parts
and use of scalar field equations of motion. We therefore can ignore them at
this order.
Similarly, the effective action
\begin{align}
  S^\text{QFT}_\text{tidal}\big|_{\text{non-linear}}= m \int d^4x \sqrt{-g}
  &\hphantom{+} \!\sum_{n=2}^\infty ( \rho_e^{(n)} \,  \phi \hat E_{\mu_1}{}^{\mu_2}\hat E_{\mu_2}{}^{\mu_3}\cdots \hat E_{\mu_n}{}^{\mu_1} \phi +\rho_m^{(n)} \, \phi \hat B_{\mu_1}{}^{\mu_2}\hat B_{\mu_2}{}^{\mu_3}\cdots \hat B_{\mu_n}{}^{\mu_1}\phi) + \cdots
\end{align}
encodes part of the lowest-multipole time-independent non-linear response,

It is not difficult to translate the different terms in the response functions into a first quantized framework. This leads to a one-to-one
relation between the higher-dimension operators in the QFT effective action and worldline operators.
The factors of $u$ are identified with the four-velocity of the worldline $u^{\mu} = dx^{\mu}/{d\tau}$ and the factors of $(u\cdot\nabla)$ 
simply become derivatives with respect to the proper time $\tau$. Thus, the analog of the operators in the effective worldline action are
\begin{align}
E_{\mu_1\mu_2\dots \mu_n}&= {\rm Sym}_{\mu_1\mu_2 \dots \mu_n}\left[P_{\mu_3}^{\nu_3}\dots P_{\mu_n}^{\nu_n} \nabla_{\nu_3}\dots \nabla_{\nu_n}
C_{\mu_1\alpha\mu_2\beta}\right] u^{\alpha}u^{\beta} \,,
\cr
B_{\mu_1\mu_2\dots \mu_n}&= {\rm Sym}_{\mu_1\mu_2\dots \mu_n}\left[P_{\mu_3}^{\nu_3}\dots P_{\mu_n}^{\nu_n} \nabla_{\nu_3}\dots \nabla_{\nu_n}
(*C)_{\mu_1\alpha\mu_2\beta}\right] u^{\alpha}u^{\beta} \,,
\cr
{E}^{(m)}_{\mu_1\dots \mu_n} &= (u^{\alpha}\nabla_\alpha)^m E_{\mu_1\dots \mu_n} = (\partial_\tau)^m E_{\mu_1\dots \mu_n} \,,
\cr
{B}^{(m)}_{\mu_1\dots \mu_n} &= (u^{\alpha}\nabla_\alpha)^m B_{\mu_1\dots \mu_n} = (\partial_\tau)^m B_{\mu_1\dots \mu_n} \,,
\end{align}
where $P_{\mu\nu}= g_{\mu\nu}-u_\mu u_\nu$ is the $u$-orthogonal projector on the worldline.
The effective action encoding the linear response are
\begin{align}
  S^\text{worldline}_\text{tidal}|_{\text{linear}}= \int d\tau
&\hphantom{+}  \sum_{n=2}^\infty \sum_{l=0}^\infty \mu^{(n,l)} \,  (E^{(l)}_{\mu_1\cdots\mu_n}E^{(l)\,\mu_1\cdots\mu_n} 
+\sigma^{(n,l)} \, B^{(l)}_{\mu_1\cdots\mu_n} B^{(l)\,\mu_1\cdots\mu_n})\,.
\end{align}
Note that here we use a different normalization than Ref.~\cite{PMTidal}, the relation between our coefficients is 
$\mu^{(n,l)}_{\text{BDG}} = 2 l! \mu^{(n,l)}$ and $\sigma^{(n,l)}_{\text{BDG}} = 2(l+1)! \sigma^{(n,l)}$.
The non-linear response is captured by
\begin{align}
  S^\text{worldline}_\text{tidal}\big|_{\text{non-linear}}=  \int d\tau
  &\hphantom{+} \sum_{n=2}^\infty  \rho_e^{(n)} \, E_{\mu_1}{}^{\mu_2}E_{\mu_2}{}^{\mu_3}\cdots E_{\mu_n}{}^{\mu_1} +\rho_m^{(n)} \, B_{\mu_1}{}^{\mu_2}B_{\mu_2}{}^{\mu_3}\cdots B_{\mu_n}{}^{\mu_1}) + \cdots \,.
\end{align}

Thus, for a particle of mass $m_i$ described by the scalar field $\phi_i$,  the correspondence between worldline  operators and QFT Lagrangian operators is
\begin{align}
\label{Esq_QFT}
&
\int d\tau E^{(l)}_{\mu_1\dots\mu_n}E^{(l)}{}^{\mu_1\dots\mu_n}
\longleftrightarrow
m_i  \int d^4 x \sqrt{-g}\phi_i {\hat E}^{(l)}_{\mu_1\dots\mu_n}{\hat E}^{(l)}{}^{\mu_1\dots\mu_n} \phi_i \,,
\\
&
\int d\tau B^{(l)}_{\mu_1\dots\mu_n}B^{(l)}{}^{\mu_1\dots\mu_n}
\longleftrightarrow
m_i \int d^4 x \sqrt{-g}\phi_i {\hat B}^{(l)}_{\mu_1\dots\mu_n}{\hat B}^{(l)}{}^{\mu_1\dots\mu_n} \phi_i 
\,.
\label{Bsq_QFT}
\end{align}
The normalization of the QFT operators is fixed such that their four-point matrix elements in the classical limit reproduce the expectation value of the worldline operators, 
provided that the normalization of the asymptotic states is the same for both of them, i.e. it is a nonrelativistic normalization for the QFT states. 
One may similarly construct a correspondence between worldline and QFT operators with more factors of the Riemann tensor. For more details about the correspondence between QFT amplitudes and worldline matrix elements see e.g. Ref.~\cite{Mogull:2020sak}.


\subsubsection{Four dimensional relations}

In any fixed dimension, the operators described above satisfy relations stemming from their finite number of components\footnote{In a different 
context these relations are known as \emph{evanescent operators} which are operators whose matrix elements vanish in
four-dimensions but not in general dimension \cite{EvanescentOperators}.};  thus they give an overcomplete description of 
the physics of extended bodies. 

One class of relations follows from the the electric and magnetic fields being
tensors of finite rank.  Naively they have rank four, but because
$E_{\mu\nu}u^{\nu}=B_{\mu\nu}u^{\nu}=0$ their rank is lowered to
three.  This is not a surprise: it is a consequence of the fact that
$E_{\mu\nu}$ and $B_{\mu\nu}$ are the covariant versions of the purely spatial
$E_{ij}, B_{ij}$ in the rest frame.  The simplest relation following from the
finiteness of the ranks of $E$ and $B$ is 
\begin{equation}
  E_{[\mu_1}{}^{\mu_2}E_{\mu_2}{}^{\mu_3}E_{\mu_3}{}^{\mu_4}E_{\mu_4]}{}^{\mu_1} = 0 \,,
\end{equation}
which, together with the tracelessness of $E$, implies that $E^4 =  1/2 (E^2)^2$. More generally, relations 
can be found which involve mixed powers of the electric and magnetic fields. For operators with no derivatives 
all such relations can be generated by evaluating the following determinant as a formal power series
\begin{equation}
  \text{det}[1+ t ( E + r B)] =  \sum_{i=2}^\infty \sum_{j=0}^i R_{i,j} t^i r^j\, .
\end{equation}
The rank-three property of an arbitrary combination of $E$ and $B$ implies that $R_{i\ge 4,j}=0$. A sample 
of such relations is
\begin{align}
  2^3R_{4,0} &=    (E^2)^2  - 2 (E^4) = 0  \,,  \nn \\
  2^2R_{4,2} &=  2 (EB)^2   +   (B^2)(E^2) - 2 (EBEB) - 4 (E^2B^2) = 0   \,, \nn \\
  5  R_{5,0} &=    (E^5)    -   \frac56 (E^2)(E^3) = 0   \,, \nn \\
  6  R_{5,2} &=  6 (E^2BEB) + 6 (E^3B^2)   - (B^2)(E^3) - 3 (E^2)(EB^2) = 0   \,, \nn \\
  2  R_{5,4} &=  2 (EB^4)   -   (B^2)(EB^2) = 0   \,,
  \label{}
\end{align}
as well as the ones that follow by interchanging $E$ and $B$. Here the round parenthesis denote the matrix trace, 
\begin{equation}
({\cal O})\equiv \Tr[{\cal O}] \,.
\label{Parentheses}
\end{equation}
Recursively solving them implies that any operator of the form $(E^{n\ge4})$ can be written as a polynomial in 
$E^2$ and $E^3$ as follows
\begin{align}
(E^n) &= n \sum_{2 p + 3 q = n}   \frac{1}{2^p 3^q}\,\frac{\Gamma(p+q)}{ \Gamma(p+1)\Gamma(q+1)} (E^2)^p\;  (E^3)^q \, .
\label{EnvsE2E3}
\end{align}
A similar relation holds for $(B^{2n})$,  while  $(B^{2n+1})=0$ in a parity-invariant theory such as GR.

Another class of relations follows from the vanishing of the Gram
determinants of any five or more four-momenta. They imply that certain
terms in the power series expansion of susceptibilities are not
linearly independent.  For instance,
\begin{equation}
  \text{det}(v_i\cdot v_j) =0 \quad \text{with} \quad  v_i \subset \{ p_1, p_2, q_1, q_2, q_3\} \,.
\end{equation}
A final class of relations, which we will not detail any further,
follows from the over-antisymmetrization of indices of both derivatives and $E$ or $B$.

An exhaustive enumeration of the $E^2$- and $B^2$-type operators was carried out in Ref.~\cite{Haddad:2020que}, 
using Hilbert series techniques~\cite{HilbertSeries}, which automatically eliminate the redundancies described here.  
In contrast, we will not make an attempt to eliminate all redundant operators, but rather use their relations 
as a check on our framework and calculations.


\section{Leading order $E^2$ and $B^2$ tidal effects}
\label{LOR2Section}

\begin{figure}[tb]
	\begin{center} \includegraphics[scale=.7]{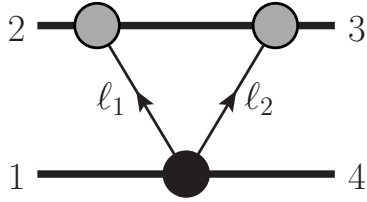} \end{center} \vskip
		-.3 cm \caption{\small The generalized cut for
                  leading-order contributions to $E^2$- or $B^2$-type
                  tidal operators.  Each blob is an on-shell
                  amplitude, which in this case is local. Each exposed
                  line is taken to be on shell and the blobs represent
                  tree amplitudes.  The dark blob contains an
                  insertion of an $E^2$- or $B^2$-type
                  higher-dimension operator with an arbitrary number
                  of additional derivatives. The external momenta are
                  all outgoing and the arrows indicated the direction of graviton momenta.  }
  \label{fig:R2cut}
\end{figure}

In this section we discuss the leading-order contribution of the
two-graviton tidal operators constructed in
Section~\ref{OperatorsSection}.  The analysis parallels to some extent
that of Ref.~\cite{Haddad:2020que}, with the main difference being the
choice of operator basis. Our choice aligns with the worldline
approach~\cite{NRGR, PMTidal} making it straightforward to compare
Love numbers. We also evaluate all integrals providing a proof of the
results with arbitrary numbers of derivatives.  Here we work in an
amplitudes-based approach following
Refs.~\cite{CliffIraMikhailClassical, 3PMPRL, 3PMLong,
  CheungSolonTidal}.

\subsection{Constructing integrands}

The first task is to write down a scattering amplitude from which
classical scattering angles and Hamiltonians can be extracted.  To
obtain the integrand we use the generalized unitarity
method~\cite{GeneralizedUnitarity}.   In this method, the integrand is
constructed from the generalized unitarity cut which we define to be
\begin{equation}
\mathcal C \equiv \sum_{\rm states} \mathcal{ M}^\tree_{(1)} \mathcal{M}^\tree_{(2)} \mathcal{M}^\tree_{(3)} \cdots
\mathcal{M}^\tree_{(m)} \, ,
\label{GenericCut}
\end{equation}
where the $\mathcal{M}^\tree_{(i)}$ are tree amplitudes, some of which
can have operator insertion.  As a simple example, \fig{fig:R2cut}
displays the unitarity cut containing the leading-order effect of an $R^2$
tidal operator.

In general, the cuts that can contribute to the conservative classical
Hamiltonian satisfy some simple rules.  The first is that generalized
unitarity cuts must separate the two matter lines to opposite sides of
a cut, which follows from the fact we are interested only in
long-range interactions.  Another general rule is that every
independent loop must have at least one cut matter line, so the energy
is restricted to a matter residue.  Any contribution with a
graviton propagator attached to the same matter line also does not
contribute to the conservative classical part.  Further details are found in
Ref.~\cite{3PMLong}.

In constructing the amplitude integrand we may immediately 
expand in soft-graviton momenta, since each power of graviton momentum 
effectively carries an additional power of $\hbar$ and is quantum 
suppressed.   This expansion can be carries out either
on at the level of the input tree amplitudes or after assembling the
cuts.  The order to which a give term needs to be expanded is dictated by
simple counting rules.  Terms with too high a scaling in the graviton momenta are
dropped.  For example, at one-loop for the case without tidal or other higher-dimension
operators this implies that any term in a diagram
numerator with more than a single power of loop momentum in the numerators yields only
quantum-mechanical contributions; some terms require fewer loop-momentum factors.
In the presence of higher-dimension operators, the leading classical 
contributions can have higher powers of loop momentum dictated simply by 
the number of extra derivatives in the operator compared to 
to the usual two derivative minimal coupling;  the extra implicit
powers of $\hbar$ are made up by the coefficient so 
the entire expression corresponds to a classical result.

In general to sew the trees together into generalized cuts one should use
physical-state projectors which depend on null reference momenta
\begin{align}
\mathbb{P}^{\mu \nu\rho\sigma} = \sum_{\rm states} \ve^{\mu\nu}(-p)
\ve^{\rho\sigma}(p) = \frac{1}{2}\Bigl( \mathbb{P}^{\mu\rho} \mathbb{P}^{\nu
	\sigma} + \mathbb{P}^{\mu\rho} \mathbb{P}^{\nu \sigma} \Bigr) - \frac{1}{D-2}
\mathbb{P}^{\mu\nu} \mathbb{P}^{\rho\sigma}\,,
\label{deDonderProjector_raw}
\end{align}
where $\mathbb{P}^{\mu\rho} = \eta^{\mu\rho} - (n^\mu p^{\rho}+n^\rho
p^{\mu})/(n\cdot p)$ and $n^\mu$ is the null reference momentum.
However, the reference momenta will drop out if the seed amplitudes
are manifestly transverse. In fact, one can always arrange for such terms to
automatically drop out~\cite{Kosmopoulos:2020pcd}.

Alternatively, we can also use four-dimensional helicity states to sew
gravitons across unitarity cuts.  In general, some caution is required
in the presence of infrared or ultraviolet singularites, although at
least through third post-Minkowskian order helicity methods have
been shown to correctly capture all contributions~\cite{3PMLong}.  For
cases without non-trivial infrared or ultraviolet
divergences~\footnote{There are ultraviolet divergence at even loop
  orders that local in momentum transfer $q$, e.g.~in the 3PM
  scattering~\cite{3PMPRL,3PMLong}. However, these are irrelevant for
  long-range dynamics because they can be absorbed by a contact
  interaction.}, we can straightforwardly apply four-dimensional
methods.
In our cases, the above $D$-dimensional sewing is simple
enough so we will not use four-dimensional helicities here.

Finally, the information from multiple generalized cuts must be merged into a
single expression.  This can either be accomplished at the level of
the integrand or after integration.  For leading tidal coefficients,
effectively only a single cut contributes, so merging information from
the cuts is trivial.

\subsubsection{Simplifications from leading classical order}
\label{LOandSpin}

The on-shell amplitudes in the unitarity cut simplifies dramatically
if we are only interested at leading classical order. Because there is
no enhancement from iteration, any terms beyond the leading order in graviton momenta
are quantum mechanical and can thus be ignored.  For example, consider a
three-point scalar-graviton-scalar amplitude at tree level
\begin{align}
{\cal M}_3(\phi(p), h(\ell),\phi(p')) = -\kappa p^\mu p^\nu \varepsilon_{\mu\nu}(\ell)\, ,
\label{threept}
\end{align}
where $\kappa$ is related to Newton's constant by $\kappa^2 = 32 \pi G$.
For any of the three-point amplitudes inserted in
Fig.~\ref{fig:R2cut}, we can replace the scalar momenta $p$ by the
external momentum $p_2$ at leading classical order.  Physically this
implies that we ignore all back reaction on the particle 2, so all
three-point amplitudes in Fig.~\ref{fig:R2cut} are approximately the
same.

For the amplitude with higher-dimension operator, it suffices to use
linearized version of the curvature operators. Expanding the metric in
the usual way, $g_{\mu\nu} = \eta_{\mu\nu}+\kappa h_{\mu\nu}$, we find
the Weyl tensor to leading order is
\begin{equation}
C_{\mu\nu \rho \sigma} =  -2\kappa \partial_{[\mu|}\partial_{[\rho}h_{\sigma]|\nu]} +{\cal O}(\kappa ^2, \Box h) \, .
\end{equation}
In deriving this expression we have also dropped terms proportional to
the equations of motion for the graviton; this is because they do not
contribute to the on-shell matrix elements necessary for the
evaluation of the leading-order amplitude.  The linearized Weyl tensor
in momentum space then reads
\begin{align}
\label{Weyl}
C^{\rm lin}_{\mu\nu \rho \sigma}(\ell) &\equiv \frac{\kappa}{2}\left[
\ell_{\mu} \ell_{\rho}\,\varepsilon(\ell)_{\nu\sigma}
-\ell_{\nu} \ell_{\rho}\,\varepsilon(\ell)_{\mu\sigma}
-\ell_{\mu} \ell_{\sigma}\,\varepsilon(\ell)_{\nu\rho}
+\ell_{\nu} \ell_{\sigma}\,\varepsilon(\ell)_{\mu\rho} \right].
\end{align}
The linearized Weyl tensor can be written a form that manifests the
double copy in terms of two gauge-theory field strengths
\begin{equation}
C^{\rm lin}_{\mu\nu \rho \sigma}(\ell) =  \frac{\kappa}{2} F^{\rm lin}_{\mu\nu}(\ell) F^{\rm lin}_{\rho \sigma} (\ell) \,,
\end{equation}
where 
\begin{equation}
F^{\rm lin}_{\mu\nu \rho \sigma}(\ell)  \equiv
\ell_{\mu} \varepsilon(\ell)_{\nu}  - \ell_{\nu} \varepsilon(\ell)_{\mu} \,,
\label{FieldStrength}
\end{equation}
and we identify the graviton polarization tensor as
$\varepsilon(\ell)_{\nu\sigma} = \varepsilon(\ell)_{\nu}
\varepsilon(\ell)_{\sigma}$.  This simple example of a double-copy
relation~\cite{KLT,BCJ}, which is trivial at the linearized level,
then implies that the leading-order amplitudes for tidal operators display double-copy relations.
The gauge invariance is manifest.

To make the gravitational coupling manifest in all equations, we will extract all factors of $\kappa$ from 
the building blocks of amplitudes. 
The linearized electric and magnetic components of the linearized Weyl tensor \eqref{Weyl} follow from Eq.~\eqref{eq:EandB_Def}
\begin{align}
&{\cal E}_{\mu_1\mu_2}(\ell, p)=
\frac{1}{2 m^2}\left[\ell_{\mu_1}\ell_{\mu_2}(p\cdot \varepsilon(\ell)\cdot p) 
-(p\cdot \ell)\left(\ell_{\mu_1}\varepsilon(\ell)_{\mu_2 \rho} p^\rho  +  \ell_{\mu_2}\varepsilon(\ell)_{\mu_1 \rho} p^\rho \right)
+ \varepsilon(\ell)_{\mu_1\mu_2}(p\cdot \ell)^2  \right],\label{calE}  \\
&{\cal B}_{\mu_1\mu_2}(\ell, p)=
\frac{1}{4 m^2}\epsilon_{\alpha\beta\gamma\mu}
\left[
(p\cdot \ell)\,(\ell^\alpha \varepsilon(\ell)^{\beta}_{\,\;\mu_2} 
-\ell^\beta \varepsilon(\ell)^{\alpha}_{\,\;\mu_2})
+ \ell^\beta \ell_{\mu_2} (p\cdot \varepsilon(\ell) )^{\alpha}
- \ell^\alpha \ell_{\mu_2} (p\cdot \varepsilon(\ell) )^{\beta}
\right],
\label{calB}
\end{align}
where the particle momentum and its four-velocity are related in the usual way, $p_\mu=m u_\mu$.
It is then straightforward to assemble the amplitude with insertions of a higher-dimension operator from above formulae.

In general to sew trees into generalized cuts one should use
physical-state projectors which depend on null reference momenta.
However, for the leading-order contributions that we will mostly be
studying here, the terms containing dependence on the reference
momentum automatically drop out because they are contracted into
manifestly gauge-invariant (transverse) quantities\footnote{In fact,
  one can always arrange for such terms to automatically drop
  out~\cite{Kosmopoulos:2020pcd}.}. Effectively, we can use the
numerator of the de Donder gauge propagator,
\begin{align}
\mathbb{P}^{\mu \nu\rho\sigma} = \sum_{\rm states} \ve^{\mu\nu}(-p)
\ve^{\rho\sigma}(p) \rightarrow \frac{1}{2}\Bigl( \eta^{\mu\rho} \eta^{\nu
  \sigma} + \eta^{\mu\rho} \eta^{\nu \sigma} \Bigr) - \frac{1}{D-2}
\eta^{\mu\nu} \eta^{\rho\sigma}\, ,
\label{deDonderProjector}
\end{align}
 to sew gravitons across cuts. 
 Combining the projector with the three-point amplitude in Eq.~\eqref{threept} at leading classical order,
effectively turns the graviton polarization tensors of the higher-dimension operator into
\begin{align}
\varepsilon_{\mu\nu}(\ell) \rightarrow  
T_{\mu\nu}(p_2)=
\left(p_{2,\mu}p_{2,\nu} - \frac{m_2^2}{D_s-2}\eta_{\mu\nu}\right) \, .
\label{eq:sewing_rule}
\end{align}
Crucially the result is independent of the loop momentum, implying that the sewing automatically imposes 
Bose symmetry for the gravitons of the higher-dimension operator. As we will outline in Sec.~\ref{LORnSection}, 
this no longer holds beyond leading order where back-reaction becomes important. For example, at next-to-leading 
order pairs of the stress tensor in Eq.~\eqref{threept} can source a single graviton, acting as a sort of ``impurity'',
which may be interpreted as the first correction to the gravitational field of a free particle towards that of a Schwarzschild black hole.

The discussion above can be extended to include the leading-order scattering of scalars deformed by higher-dimension operators off 
higher-spin particles described the Lagrangian in Ref.~\cite{Bern:2020buy}. For a generic spinning body the stress tensor is
\begin{align}
\label{Lstresstensor}
{\cal M}_3(\phi_s(p), h(\ell),\phi_s(p')) &= - \kappa \,V_3^{\mu\nu}(\phi_s(p), h(\ell),\phi_s(p'))\varepsilon_{\mu\nu}(\ell) \,,
\\
V_3^{\mu\nu}(\phi_s(p), h(\ell),\phi_s(p')) &=
{p^{\mu}p^{\nu}}
\sum_{n=0}^{\infty} \frac{C_{ES^{2n}}}{\left(2n\right)!}
\left(\frac{\ell\cdot S(p)}{m}\right)^{2n}  \!\!
-i \ell_\rho p^{(\mu}S(p)^{\nu)\rho}
\sum_{n=0}^{\infty} \frac{C_{BS^{2n+1}}}{\left(2n+1\right)!}
\left(\frac{\ell\cdot S(p)}{m}\right)^{2n} \! ,
\nonumber
\end{align}
where $\ell$ is the graviton momentum and $S(p)^\mu$ and $S(p)^{\mu\nu}$ are the covariant spin vector and spin tensor, related by
\begin{align}
S^{\mu\nu}(p)= -\frac{1}{m}\epsilon^{\mu\nu\gamma\delta}{p}_{\gamma}{S}_{\delta}(p)
\,, \hskip 2 cm 
S^\mu(p) = -\frac{1}{2m} \epsilon^{\mu\beta\gamma\delta}{p}_{\beta}{S}_{\gamma\delta}(p) \, ,
\end{align}
and we recall that in the classical limit $\ell\cdot S(p)/m = {\cal O}(1)$.

For the Kerr black hole the stress tensor, originally found in Ref.~\cite{Vines:2017hyw} from different considerations, is obtained by setting 
$C_{ES^{2n}}=C_{BS^{2n}}=1$ and has the closed-form expression
\begin{align}
{\cal M}^\text{Kerr}_3(\phi_s(p), h(\ell),\phi_s(p')) = -\kappa \exp(ia*\ell){}^{(\mu}{}_\rho p^{\nu)}p^\rho \varepsilon_{\mu\nu}(\ell) \, ,
\end{align}
where 
\begin{align}
a^\mu = \frac{1}{2p^2} \epsilon{}^\mu{}_{\nu\rho\sigma}p^\nu S^{\rho\sigma}(p) \,,
\hskip 1.5 cm 
(a*\ell)^\mu{}_{\nu} \equiv \epsilon^{\mu}{}_{\nu\rho\sigma}a^{\rho}\ell^\sigma \, .
\label{adef}
\end{align}

Despite the more complicated dependence on the graviton momentum, the sewing of the spinning three-point amplitudes 
with the composite operator contact term can be carried by a replacement analogous to Eq.~\eqref{eq:sewing_rule}. For example,  
for a particle with the stress of a Kerr black hole, it is
\begin{align}
\label{Kerr_sewing}
\varepsilon_{\mu\nu}(\ell) \rightarrow  
T^\text{Kerr}_{\mu\nu}(\ell, p_2)=
\exp(ia*\ell){}^{(\alpha}{}_\rho p^{\beta)}p^\rho\left(\delta_\alpha^\mu\delta_\beta^\nu - \eta_{\alpha\beta} \frac{\eta^{\mu\nu}}{D_s-2}\right) \, .
\end{align}
We note that only the terms with an even number of spin vectors, in general governed by the coefficients $C_{ES^{2n}}$, contribute
to the trace part of this replacement.  
To shorten the ensuing equations, in the following we will use the replacement
\begin{align}
\label{sewingwithspin}
\epsilon^{\mu\nu}(\ell) \rightarrow T_\text{gen}^{\mu\nu}(\ell, p_2) = 
\left(p_2^\mu p_2^\nu-  \frac{m_2^2}{D_s - 2} \eta^{\mu\nu}\right) A(\ell) 
 -\frac{i}{2} 
 \ell_\rho (p_2^\mu S^{\nu\rho}(p_2) + p_2^\nu S^{\mu\rho}(p_2) ) B(\ell) \,,
 \end{align}
where $A(\ell)$ and $B(\ell)$ can be read off Eqs.~\eqref{Lstresstensor} and \eqref{Kerr_sewing}. 

\subsection{Momentum-space analysis}
\label{1loop_eg_p_space}

Before discussing the leading-order effects of the most general tidal operators introduced in Sec.~\ref{OperatorsSection},
we discuss here the simpler case of operators $E_{\mu_1\mu_2}^{(m)}$, corresponding to the multipoles of the gravitational 
field of the quadrupole operator $E_{\mu\nu}$. 

The construction of the relevant four-point matrix element of the operator $\phi { E}_{\mu_1\mu_2}^{(m)} { E}^{(m)}{}^{\mu_1\mu_2}\phi$, 
corresponding to the darker blob in Fig.~\ref{fig:R2cut}, is straightforward.
The matrix element is
\begin{align}
\label{M4Esq}
{\cal M}_{{\rm E}^2_{l,2}}(h(\ell_1), h(\ell_2), \phi(p_1), \phi(p_4) ) 
&=
2\kappa^2 m_1\left(D_{{\rm E}^2_{l,2}}(p_1, \ell_1, p_4, \ell_2)+D_{{\rm E}^2_{l,2}}(p_1, \ell_2, p_4, \ell_1) \right) ,
\nn \\
D_{{\rm E}^2_{l,2}}(p_1, \ell_1, p_4, \ell_2)&=\left(\frac{i}{m_1}\right)^{2l}(p_1\cdot \ell_1)^{l}(p_1\cdot \ell_2)^{l} 
 {\cal E}_{\mu_1\mu_2}(\ell_1, p_1) {\cal E}^{\mu_1\mu_2}(\ell_2, p_4)\,.
\end{align}
As noted earlier, because tidal operators are gauge invariant and constructed out of Weyl tensors, this matrix element obeys 
the transversality conditions for the two gravitons. Thus, their contribution to generalized unitarity cut in Fig.~\ref{fig:R2cut} automatically 
accounts for the physical-state projection. 
The sewing is then simply given by the replacement in Eq.~\eqref{eq:sewing_rule}. To leading order in soft expansion we can also 
replace all $p_1\cdot \ell_2 = -p_1\cdot \ell_1 +\mathcal{O}(q)$.

The resulting amplitude is
\begin{align}
\label{Esqsewing}
{\cal M}_{{\rm E}^2_{l,2}} (\bm p,\bm q)
&= i \kappa^2\,  \int \frac{d^D \ell_1}{(2\pi)^D} \frac{{\cal M}_{{\rm E}^2_{l,2}} (h(\ell_1), h(\ell_2), \phi(p_3), \phi(p_4))\big\vert_{\varepsilon_{\mu\nu}(\ell_i) \rightarrow  
		T_{\mu\nu}(p_2)}}{\ell_1^2 ((\ell_1-p_2)^2-m_2^2) (q-\ell_1)^2} \nn \\
&= 4i \, m_1\kappa^4
\, \int \frac{d^D \ell_1}{(2\pi)^D} \frac{(u_1\cdot \ell_1)^{2l} {\cal E}_{\mu_1\mu_2}(\ell_1, p_1) {\cal E}^{\mu_1\mu_2}(\ell_2, p_1)\big\vert_{\varepsilon_{\mu\nu}(\ell_i) \rightarrow  
	T_{\mu\nu}(p_2)}}
{\ell_1^2 ((\ell_1-p_2)^2-m_2^2) (q-\ell_1)^2} \,,
\end{align}
where the numerator is given more explicitly by
\begin{align}
\label{EsqEval}
&{\cal E}_{\mu_1\mu_2}(\ell_1, p_1) {\cal E}^{\mu_1\mu_2}(\ell_2, p_1)\big\vert_{\varepsilon_{\mu\nu}(\ell_i) \rightarrow  
	T_{\mu\nu}(p_2)}   \\
& \quad\; =\frac{1}{8}
 m_2^4\left[ (u_1\cdot \ell_1)^2((u_1\cdot \ell_1)^2  + {\frac{1}{2}} q^2)
-2 \sigma^2 q^2 (u_1\cdot \ell_1)^2
+\frac{1}{8}  q^4(1 - 2 \sigma^2)^2\right]+{\cal O}(q^6) \, . \nonumber
 \end{align}
Further expanding the amplitude in the soft limit leads to
\begin{align}
{\cal M}_{{\rm E}^2_{l,2}} (\bm p, \bm q)   =  64 i \pi^2 G^2  |\bm q|^{3 + 2 l} m_1  m_2^3 ((1 - 2 \sigma^2)^2 I_{2l} + 
   4 (-1 + 4 \sigma^2) I_{2(1 + l)} + 8 I_{2(2 + l)}) \,,
   \nonumber
\end{align}
where $I_{2l}$ are triangle integrals
\begin{align}
\label{triangles0}
I_{2l} = \int\frac{d^D \ell}{(2\pi)^D} \frac{|\bm{q}|^{-2l+1}(\ell\cdot u_1)^{2l}}{ \ell^2 ( -2 \ell\cdot u_2) (\ell - q)^2 }  \,,
\end{align}
which must be evaluated in the potential region. The results of these integrals were conjectured in Ref.~\cite{Haddad:2020que}. 
Here we present the proof, by going to the frame in which particle~2 is at rest
\begin{equation}
  \label{eq:restframe2}
  u_{1\mu} =  -(\sigma,0,0,\sqrt{\sigma^2-1}) \,, \qquad u_{2\mu} =-(1,0,0,0)\,,
  \qquad q_\mu = (0,\bm q) = (0,q^x,q^y,q^z) \,,
\end{equation}
under which $\ell_{i\mu} = (\ell^0_i,\bm \ell_i) = (\ell^0_i,\ell^x_i,\ell^y_i,\ell^z_i)$. Note that since $q^z = \bm q \cdot \hat{\bm z} = \mathcal{O}(q^2)$ by on shell conditions, we can treat $q^z \approx0$ if we are only interested in the leading classical limit.
We then have
\begin{equation}
  I_{2l}=(\sigma^2-1)^{l} \int \frac{d^D\ell}{(2\pi)^{D}} \frac{ |\bm{q}|^{-2l+1} (\ell^z)^{2l}}{(2\ell^0)\ell^2(\ell-q)^2} =  \frac{i(\sigma^2-1)^{l}}{2^{2l+1}(4\pi)^{(D-1)/2}} \int \frac{d^{D-1}\bm\ell}{\pi^{(D-1)/2}} \frac{|\bm{q}|^{-2l+1}( 2\ell^z)^{2l}}{\bm\ell^2(\bm\ell-\bm q)^2} \,,
\end{equation}
where in the second equality we have evaluated the residue of the energy pole with a symmetry factor $1/2$ because the graviton propagators cannot be on shell in the potential region.
The remaining integral is a Euclidean triangle with a linearized propagator and is given by Smirnov in Ref.~\cite{SmirnovBook},
\begin{align}
&\int\!\frac{\mathrm{d}^{D-1}\bm{\ell}}{\pi^{(D-1)/2}}\frac{(\bm{q}^2)^{a+b+\frac{c}{2}-\frac{3}{2}}}{(\bm{\ell}^2-i 0)^a[(\bm{\ell}-\bm{q})^2-i 0]^b(2\ell^z - i 0 )^c}
\label{eq:SmirnovMasterTriangle}\\
&\hspace{2cm} ={}e^{\frac {i \pi c} {2}} |\bm{q}|^{-2\epsilon} \frac{\Gamma \left(\frac{c}{2}\right) \Gamma \left(\frac{3}{2}-a-\frac{c}{2}-\epsilon\right) \Gamma \left(\frac{3}{2}-b-\frac{c}{2}-\epsilon \right)
	\Gamma \left(a+b+\frac{c}{2}+\epsilon -\frac{3}{2}\right)}{2\Gamma (a) \Gamma (b) \Gamma (c) \Gamma (3-a-b-c-2\epsilon)} \,, \nonumber
\end{align}
for $\bm q\cdot \hat{\bm z} =0$ which is valid for leading order in the classical limit.
The result is 
\begin{equation}
  I_{2l} = -\frac{i(\sigma^2 - 1)^{l}}{4^{l+2-\epsilon}(4\pi)^{1/2-\epsilon}} |\bm{q}|^{-2\epsilon} \frac{\Gamma \left(\frac{1}{2}-\epsilon\right) 
\Gamma \left(\frac{1}{2}+\epsilon \right)}{\Gamma\left(\frac{1}{2}-l\right) 
  \Gamma\left(1-\epsilon + l\right)} \,.
\label{eq:oneloopint}
\end{equation}
Using the result for these integrals with $\epsilon = 0$ the amplitude is 
\begin{align}
{\cal M}_{{\rm E}^2_{l,2}}(\bm p, \bm q) 
&= |\bm q|^{3+2l} {\overline{\cal M}}_{{\rm E}^2_{l,2}}(\bm p) \,,
\label{AmplitudeEneq2m} \\
{\overline{\cal M}}_{{\rm E}^2_{l,2}}(\bm p) &=
G^2m_1 m_2^3 \, \frac{(-1)^l\pi^{3/2}\Gamma(\textstyle{\frac{1}{2}} + l)}{2^{2 (1 + l)}\Gamma(3 + l)}  
\\
&\hspace{.5cm}\times 
(\sigma^2-1)^l  (11 + 4 l (3 + l) -6(5 + 2 l) \sigma^2 + (5 + 2 l) (7 + 2 l) \sigma^4)  \,.
\nonumber
\end{align}
%
The corresponding potential and eikonal phase are
\begin{align}
\label{PotentialEneq2m}
V_{{\rm E}^2_{l,2}}(\bm p, \bm r) &=   \frac{-1}{4 E_1 E_2 |\bm r |^{2l+6}}   \,  \frac{2^{3+2 l}\Gamma(3+l)}{\pi^{3/2} \Gamma(-\textstyle{\frac{3}{2}} - l)} 
  \,      {\overline{\cal M}}_{{\rm E}^2_{l,2}}(\bm p)
 \,,
\\
\delta_{{\rm E}^2_{l,2}}(\bm p, {\bm b})  &=  \frac{1}{4m_1 m_2\sqrt{\sigma^2-1}}\,   \frac{1}{ |\bm b |^{2l+5}}   \,  
\frac{2^{3+2 l}\Gamma(\textstyle{\frac{5}{2}} + l)}{\pi \Gamma(-\textstyle{\frac{3}{2}} - l)} 
  \, {\overline{\cal M}}_{{\rm E}^2_{l,2}}(\bm p)
\,.
\label{eikphasen=2m}
\end{align}
%
It is not difficult to see that, for $l=0$ and $l=1$, eq.~\eqref{eikphasen=2m} reproduces the expectation values of the operators 
$E^2$ and $({\dot E})^2$ evaluated in Ref.~\cite{PMTidal}.

The calculation above can be easily repeated for the operator $B_{\mu\nu}^{(l)}B^{\mu\nu}{}^{(l)}$; it amounts to replacing in 
Eq.~\eqref{Esqsewing} ${\cal E}$ with ${\cal B}$ given in Eq.~\eqref{calB}.
The resulting amplitude, potential and eikonal phase are:
\begin{align}
&{\cal M}_{{\rm B}^2_{l,2}}(\bm p, \bm q) 
= |\bm q|^{3+2l} {\overline{\cal M}}_{{\rm B}^2_{l,2}}(\bm p) \,,
\label{AmplitudeBneq2m} \\
&{\overline{\cal M}}_{{\rm B}^2_{l,2}}(\bm p) =
G^2 m_1 m_2^2 \frac{(-1)^l\pi^{3/2}    \Gamma(\textstyle{\frac{1}{2}} + l)   }{2^{2(l+1)} \Gamma(3 + l)}  (5 + 2 l) 
 (\sigma^2-1)^{l+1} (1 + 2 l + (7 + 2 l) \sigma^2) \,,
\\
&V_{{\rm B}^2_{l,2}}(\bm p, \bm r) =   \frac{-1}{4 E_1 E_2 |\bm r |^{2l+6}}   \,  \frac{2^{3+2 l}\Gamma(3+l)}{\pi^{3/2} \Gamma(-\textstyle{\frac{3}{2}} - l)} 
  \,      {\overline{\cal M}}_{{\rm B}^2_{l,2}}(\bm p) \,,
\\
&\delta_{{\rm B}^2_{l,2}}(\bm p, {\bm b})  =  \frac{1}{4m_1 m_2\sqrt{\sigma^2-1}}\,   \frac{1}{ |\bm b |^{2l+5}}   \,  
\frac{2^{3+2 l}\Gamma(\textstyle{\frac{5}{2}} + l)}{\pi \Gamma(-\textstyle{\frac{3}{2}} - l)} 
  \, {\overline{\cal M}}_{{\rm B}^2_{l,2}}(\bm p)
\,.
\end{align}
Similarly to eq.~\eqref{eikphasen=2m}, the eikonal phase above evaluated on 
$l=0$ and $l=1$ reproduces the expectation values of the operators $B^2$ and $({\dot B})^2$ found in~\cite{PMTidal}.

\subsection{Position-space analysis}
\label{real_oneloop}
Alternatively, the calculation can be done in position space, more specifically in the rest frame of particle 2 as in Eq.~\eqref{eq:restframe2}. 
This approach will provide a simple way to generalize the analysis beyond one loop.
There are two key observations here.  First, the amplitude with $C^2$
operator insertion in Eq.~\eqref{M4Esq} factorizes into a product of
the multipole expansions of electric or magnetic tensors
\begin{align}
{\cal M}_{{\rm E}^2_{l,2}}(h(\ell_1), h(\ell_2), \phi(p_1), \phi(p_4) )
&=
4m_1 \kappa^2 \left(\frac{i}{m_1}\right)^{2l}
((p_1\cdot \ell_1)^{l} {\cal E}_{\mu_1\mu_2}(\ell_1, p_1))
((p_1\cdot \ell_2)^{l} {\cal E}^{\mu_1\mu_2}(\ell_2, p_1)) \nn \\[3pt]
& \hskip 2 cm \null  +\mathcal{O}(q^{2l+4})\,,
\label{eq:fact_example}
\end{align}
where we have applied the classical limit $p_4 = -p_1 +\mathcal{O}(q)$ to Eq.~\eqref{M4Esq}.
Second, 
in the potential region, we can integrated out graviton energy component by picking up residue from the matter propagator~\cite{CliffIraMikhailClassical,3PMLong}.
This sets $\ell^0_1 = \ell^0_2 = 0$ and implies the graviton momenta $\ell_1,\ell_2$ are purely spatial. To exploit the factorization at the integrand level, we further Fourier transform the spatial $\bm q$ in Eq.~\eqref{Esqsewing} to position space\footnote{The Fourier transform acts on the amplitude with generic off-shell $q$, which is three dimensional. We use $\widetilde{\mathcal{M}}(\bm p,\bm q)$ to denote amplitude with off-shell $\bm q$.}
\begin{align}
\mathcal{M}_{{\rm E}^2_{l,2}}(\bm p,\bm r) 
\label{positionspace_1loop}
&\equiv \int \frac{d^{D-1}\bm q}{(2\pi)^{D-1}} e^{-i\bm r \cdot \bm q}\, \widetilde{\mathcal{M}}_{{\rm E}^2_{l,2}} (\bm p,\bm q)  \\
&= 
\frac{\kappa^2}{4m_2}\,
\prod_{i=1}^{2}\int \frac{d^{D-1}\bm \ell_i}{(2\pi)^{D-1}}\, \frac{e^{-i\bm r\cdot \bm \ell_i}}{\bm \ell_i^2}\,
{\cal M}_{{\rm E}^2_{l,2}}(h(\ell_1), h(\ell_2), \phi(p_1), \phi(p_4) )\big\vert_{\varepsilon_{\mu\nu}(\ell_i) \rightarrow  
	T_{\mu\nu}(p_2)} .\nn
\end{align}
Crucially, the dependence on the two graviton momenta $\ell_1,\ell_2$ factorizes and each of them can be treated as an independent variable. 
Together with the factorization in Eq.~\eqref{eq:fact_example}, the Fourier transform acts on individual electric tensor 
${\cal E}_{\mu_1\mu_2}(\ell_i, p_1)$.  We define
\begin{align}
\label{massagedE}
{\cal E}_{\mu \nu}(\bm r, p_1) \equiv &
\int \frac{d^{D-1}\bm \ell_i}{(2\pi)^{D-1}} \frac{e^{-i\bm r\cdot \bm \ell_i}}{\bm \ell_i^2}
{\cal E}_{\mu \nu}(\ell_i, p_1)\big\vert_{\varepsilon_{\rho\sigma}(\ell) \rightarrow  
	T_{\rho\sigma}(p_2)} \nn \\
=&
\frac{-m_2^2}{16\pi |\bm r|^5} \Big[
3\left(\bm r^2 +2 (\sigma^2-1)z^2 \right) u_{2\mu}u_{2\nu}
-3\sigma \bm r^2 (u_{2\mu} u_{1\nu}+u_{2\mu}u_{1\nu})
+ 2\bm r^2 u_{1\mu}u_{1\nu}
 \nn \\
& \hskip 1.2 cm 
+3 (2\sigma^2-1) r_{\mu} r_{\nu}
-6\sigma \sqrt{\sigma^2-1} z \, (u_{2\mu}r_{\nu}+u_{2\mu}r_{\nu})
+3 \sqrt{\sigma^2-1} z \, (u_{1\mu}r_{\nu}+u_{1\mu}r_{\nu}) \nn \\
&\hskip 1.2 cm 
+ ((3\sigma^2-2)\bm r^2 -3(\sigma^2-1) z^2) \eta_{\mu\nu}
\Big],
\end{align}
where $r_\mu = (0,\bm r)=(0,x,y,z)$ in the frame of Eq.~\eqref{eq:restframe2} 
as the electric field sourced by $p_2$ in position space.
The Fourier transform of 
scalar-graviton amplitude (with the graviton propagators) is then
\begin{align}
&\, {\cal M}_{{\rm E}^2_{l,2}}\left(h_1, h_2, \phi(p_1), \phi(p_4)|\bm r\right) \equiv 
\prod_{i=1}^2 \int \frac{d^{D-1}\bm \ell_i}{(2\Pi)^{D-1}} \frac{e^{-i\bm r\cdot \bm \ell_i}}{\bm\ell_i^2}
{\cal M}_{{\rm E}^2_{l,2}}(h(\ell_1), h(\ell_2), \phi(p_1), \phi(p_4) )
\nn \\
 & \qquad = \,
{\cal M}_{{\rm E}^2_{l,2}}\left(h(\ell_1), h(\ell_2), \phi(p_1), \phi(p_4)\vert 
{\cal E}_{\mu_1\mu_2}(\ell_j, p_1) \rightarrow {\cal E}_{\mu_1\mu_2}(\bm r_j, p_1),
\ell_i \rightarrow i\nabla_j
\right)\big \vert_{\bm r_j \rightarrow \bm r},
\label{eq:position_compton}
\end{align}
where any loop momentum $\ell_j$ is replaced with the gradient on the position $\bm r_j$ 
of the electric field ${\cal E}_{\mu_1\mu_2}(\bm r_j, p_1)$
and all $\bm r_j$ are identified with $\bm r$.
The two-scalar scattering amplitude in position space then has a simple form
\begin{align}
\mathcal{M}_{{\rm E}^2_{l,2}}(\bm p,\bm r) 
&= \frac{\kappa^2}{4m_2}\,
{\cal M}_{{\rm E}^2_{l,2}}\left(h_1, h_2, \phi(p_1), \phi(p_4)|\bm r\right) \nn \\
&= \kappa^4\frac{m_1}{m_2}\, (\sigma^2-1)^{l}
\left[
(\hat{\bm z}\cdot\nabla)^{l}\, {\cal E}_{\mu_1\mu_2}(\bm r, p_1)
\right]^2 \,,
\end{align}
where in the second line we plug in the result in Eq.~\eqref{eq:fact_example}, apply
the replacement in Eq.~\eqref{eq:position_compton} and $\hat{\bm z}$ is the unit vector along $z$ direction.

The position-space result is generally not isotropic; namely, it could depend on $\hat{\bm z} \cdot \bm r$.
To make the result isotropic, we go back to momentum space and impose the on-shell condition $\hat{\bm z}\cdot \bm q =\mathcal{O}(\bm q^2) \simeq 0$,
\begin{align}
\mathcal{M}_{\mathcal{O}}(\bm p,\bm q) = \int  d^{D-1}\bm r \, e^{+i\bm r\cdot \bm q}\, \mathcal{M}_{\mathcal{O}}(\bm p, \bm r) \big\vert_{\hat{\bm z}\cdot \bm q =0} \,.
\label{eq:fourierToMom}
\end{align}
Since the result only depends on the covariant variables $\sigma$ and $q^2=-\bm q^2$, it can be promoted to any other frame.
All Fourier-transforms that appear in this calculation are of the form
\begin{align}
\int {d^{D-1} \bm r} \frac{e^{i \bm r \cdot \bm q} ({\hat{\bm z}}\cdot \bm r)^s}{\bm r^{h}}
= \frac{(-1)^{s/2} \pi^{D/2}}{2^{h-s-D+1}} \frac{ |\bm q|^{h-s-D+1}}{\sin (\textstyle{\frac{1}{2}}\pi (D-1-h))} 
\frac{\Gamma(\textstyle{\frac{1}{2}}(1+s))}{\Gamma(\textstyle{\frac{1}{2}}h)\Gamma(1+\textstyle{\frac{1}{2}}(h-s-D+1))} \ ,
\label{eq:FT_generalTensor}
\end{align}
for some exponents $h$ and integer $s$. The isotropic potential then follows from Eq.~\eqref{eq:VMbar}.

From the position-space amplitude we can directly obtain the eikonal phase, although it can be calculated easily once we have the amplitude ${\cal M}_{\mathcal{O}}(\bm p, \bm b)$.
To see this, we simply invert the amplitude in terms of  Eq.~\eqref{VvsM} and plug it into Eq.~\eqref{PhasevsM}
\begin{align}
\delta_{\mathcal{O}}(\bm p, \bm b) &= \frac{1}{4m_1 m_2 \sqrt{\sigma^2-1}} 
\int  \frac{d^{D-2}\bm q}{(2\pi)^{D-2}} e^{-i\bm b\cdot \bm q} 
\int  d^{D-1}\bm r e^{i\bm r\cdot \bm q} \,
{\cal M}_{\mathcal{O}}(\bm p,\bm r)\Big \vert_{\bm q = (q^x,q^y,0)} \nn \\
&= \frac{1}{4m_1 m_2 \sqrt{\sigma^2-1}}\,
\int^{\infty}_{-\infty}  dz \, {\cal M}_{\mathcal{O}}(\bm p,\bm r=(\bm b, z)),
\end{align}
where we use $\bm b = (b^x,b^y,0)$ and $\bm r = (x,y,z)$. Since we are only interested in the leading order, the particle trajectory can be treated as a straight line. In the frame where particle 2 is rest at the origin, the position of particle 1 is $x^\mu_1 =(t,\bm r)= b^\mu + u_1^\mu \tau =\tau(\sigma,b^x,b^y,\sqrt{\sigma^2-1})$. The above formula can be written as
\begin{align}
\label{eq:phaseFromPosition}
\delta_{\mathcal{O}}(\bm p, \bm b) 
&= \frac{1}{4m_1 m_2}\,
\int^{\infty}_{-\infty}  d\tau \, {\cal M}_{\mathcal{O}}(\bm p,\bm r(\tau))\,.
\end{align}
So the eikonal phase can be obtained straightforwardly from ${\cal M}_{\mathcal{O}}(\bm p,\bm r(\tau))$. This is expected because the eikonal phase is proportional to the worldline action integrated over a straight line.
Our approach here offers a derivation from purely scattering-amplitudes perspective.

The advantage of position-space approach is that it is very general. The discussion above applies to contribution of any tidal
operator at its leading classical order. The only integrals needed, to any loop order, are in Eq.~\eqref{eq:FT_generalTensor}.
We will discuss and illustrate this point in more detail in Sec.~\ref{LORnSection}.

The discussion above can be generalized easily to the case with magnetic operators. The position-space magnetic component 
of the linearized Weyl tensor, contracted with a point-particle stress tensor, is
\begin{align}
\label{massagedB}
{\cal B}_{\mu \nu}(\bm r, p_1) \equiv &
\int \frac{d^{D-1}\bm \ell_i}{(2\pi)^{D-1}} \frac{e^{-i\bm r\cdot \bm \ell_i}}{\bm \ell_i^2}
{\cal B}_{\mu \nu}(\ell_i, p_1)\big\vert_{\varepsilon_{\rho\sigma}(\ell) \rightarrow  
	T_{\rho\sigma}(p_2)}.
\end{align}
We have the scalar-graviton amplitude in position space
\begin{align}
&\, {\cal M}_{{\rm B}^2_{l,2}}\left(h(\ell_1), h(\ell_2), \phi(p_1), \phi(p_4)|\bm r\right) \equiv 
\prod_{i=1}^2 \int \frac{d^{D-1}\bm \ell_i}{(2\Pi)^{D-1}} \frac{e^{-i\bm r\cdot \bm \ell_i}}{\bm\ell_i^2}
{\cal M}_{{\rm B}^2_{l,2}}(h(\ell_1), h(\ell_2), \phi(p_1), \phi(p_4) )
\nn \\
 & \qquad = 
{\cal M}_{{\rm B}^2_{l,2}}\left(h(\ell_1), h(\ell_2), \phi(p_1), \phi(p_4)\vert 
{\cal B}_{\mu_1\mu_2}(\ell_j, p_1) \rightarrow {\cal B}_{\mu_1\mu_2}(\bm r_j, p_1),
\ell_i \rightarrow i\nabla_j
\right)\big \vert_{\bm r_j \rightarrow \bm r}.
\label{eq:position_compton_B}
\end{align}
Again we identify all $\bm r_j$ in the end with $\bm r$. The position-space amplitude is then
\begin{align}
\mathcal{M}_{{\rm B}^2_{l,2}}(\bm r) 
&= \frac{1}{m_2}\left(\frac{\kappa}{2}\right)^2\,
{\cal M}_{{\rm B}^2_{l,2}}\left(h_1, h_2, \phi(p_1), \phi(p_4)|\bm r\right) .
\end{align}

Let us comment on an interesting relation between electric and magnetic operators. In position space we find
\begin{align}
{\cal E}_{\mu \nu}(\bm r, p_1){\cal E}^{\mu \nu}(\bm r, p_1)
&=
\frac{3m_2^4}{128\pi |\bm r|^{10}}\, \left[3(\sigma^2-1)(\bm r^2- z^2)(\sigma^2 \bm r^2 -(\sigma^2-1)z^2)+\bm r^4
\right] ,
\label{eq:ESq_real}
\\
{\cal B}_{\mu \nu}(\bm r, p_1){\cal B}^{\mu \nu}(\bm r, p_1)
&=\frac{9m_2^4}{128\pi |\bm r|^{10}}\, (\sigma^2-1)(\bm r^2- z^2)(\sigma^2 \bm r^2 -(\sigma^2-1)z^2) \,.
\label{eq:BSq_real}
\end{align}
The two operators are almost identical. The difference between the two
is independent of $\sigma$ which is sub-sub-leading in the high-energy
limit $\sigma \gg 1$. As explained in Ref.~\cite{PMTidal}, this is
expected because the difference is proportional to Weyl tensor squared
which is independent of $\sigma$. This behavior has also been observed
at the next-to-leading order in Ref.~\cite{PortoTidalTwoLoop}.

\subsection{General multipole operators}

Following the example discussed in detail in the previous sections, we proceed to evaluate the amplitudes and the corresponding eikonal 
phases with one insertion of the generic tidal operators $\phi { E}{ }^{(l)}_{\mu_1\dots\mu_n} { E}{}^{(l)}{}^{\mu_1\dots\mu_n}\phi$ 
and $\phi B^{(l)}_{\mu_1\dots\mu_n}  B^{(l)}{}^{\mu_1\dots\mu_n}\phi$. As already mentioned for operators with $n=2$, 
we may choose without loss of generality, the two $ E$ and $ B$ factors to have equal upper index.

The calculations for the two operators are parallel. For this reason, in the common part we will collectively denote $E$ or $B$ by $X$,
and specialize at them at the end. Thus, to leading order in $\kappa$, the momentum space expressions of 
${ {\hat E}}{}^{(l)}$ and ${ {\hat B}}{}^{(l)}$  defined in Eq.~\eqref{QFToperators} are
\begin{align}
\label{momentumspaceE}
 { { X}}{}^{(l)}_{\mu_1\mu_2\dots\mu_n} = i^{2l + (n-2)}
\left(\frac{i}{m}\right)^{l}
 (p\cdot \ell)^l 
{\rm Sym}_{\mu_1\dots \mu_n}[ P_{\mu_3}^{\nu_3}(p) \ell_{\nu_3}\dots P_{\mu_n}^{\nu_n}(p) \ell_{\nu_n}{X}(\ell, p)_{\mu_1\mu_2}] + {\cal O}(\kappa^2) \,,
\end{align}
where $P_{\mu_i}^{\nu_i}$ are the momentum space form of the projectors in Eq.~\eqref{u_projector} and 
$X_{\mu_1\mu_2}(\ell, p)$ being given by ${\cal E}_{\mu_1\mu_2}$ and ${\cal B}_{\mu_1\mu_2}$ in Eqs.~\eqref{calE}-\eqref{calB} for 
the two operators, respectively.
The symmetrization over the indices $\mu_1,\dots, \mu_n$ includes division by the number of terms. In the expression above 
 $\ell$ is the graviton momentum, $p$ is the scalar momentum and $\varepsilon(\ell)$ in the explicit expressions of ${\cal E}_{\mu_1\mu_2}$ 
 and ${\cal B}_{\mu_1\mu_2}$ is the graviton polarization tensor. 

The product of two linearized $X^{(l)}_{\mu_1\dots \mu_n}$ with different graviton momenta 
$\ell_1$ and $\ell_2$,  and contracted as in Eqs.~\eqref{Esq_QFT} and \eqref{Bsq_QFT},  contains three different structures:
(1) all projectors are contracted with each other, 
(2) all but one projector are contracted with each other and
(3) all but two projectors are contracted with each other.
The four-point matrix element of  the operator $ \phi { X}^{(l)}_{\mu_1\dots\mu_n}{ X}^{(l)}{}^{\mu_1\dots\mu_n}\phi$ 
needed for the construction of the four-scalar amplitude is
\begin{align}
{\cal M}_{{\rm X}^2_{l,n}}(h(\ell_1), h(\ell_2), \phi(p_1), \phi(p_4) )
&=
2\kappa^2 m_1\left(D_{{\rm X}^2_{l,n}}(p_1, \ell_1, p_4, \ell_2)+D_{{\rm X}^2_{l,n}}(p_1, \ell_2, p_4, \ell_1) \right)  ,
\end{align}
where
\begin{align}
D_{{\rm X}^2_{l,n}}(p_1, \ell_1, p_4, \ell_2)&=i^{2(n-2)}
i^{2l}(-1)^l
\frac{2 (n-2)!}{n!}(u_1\cdot \ell_1)^{2l} \Big[ 
 (\ell_1\cdot P(p_1)\cdot P(p_4)\cdot \ell_2)^{n-2} \Pi^{X}_1(p_1, \ell_1, p_4, \ell_2)
\nn    \\
& \qquad  \qquad 
+ 2 (n-2)(\ell_1\cdot P(p_1)\cdot P(p_4)\cdot \ell_2)^{n-3} \Pi^{X}_2 (p_1, \ell_1, p_4, \ell_2)
   \\
& \qquad  \qquad 
+  \frac{1}{2}(n-2)(n-3)(\ell_1\cdot P(p_1)\cdot P(p_4)\cdot \ell_2)^{n-4} \Pi^{X}_3(p_1, \ell_1, p_4, \ell_2) \Big] \, .
\nonumber
\end{align}
The three factors $\Pi^{X}_1(p_1, \ell_1, p_4, \ell_2)$ are given by
\begin{align}
\label{PiE}
\Pi^{X}_1(p_1, \ell_1, p_4, \ell_2) &= X_{\mu_1\mu_2}(\ell_1, p_1) X^{\mu_1\mu_2}(\ell_2, p_4) \,,
\\
\Pi^{X}_2(p_1, \ell_1, p_4, \ell_2) &= \ell_1\cdot P(p_1)\cdot X(\ell_2, p_4)\cdot X(\ell_1, p_1)\cdot P(p_4)\cdot \ell_2 \,,
\cr 
\Pi^{X}_3(p_1, \ell_1, p_4, \ell_2) &= 
\ell_1\cdot P(p_1)\cdot X(\ell_2, p_4)\cdot P(p_1) \cdot \ell_1 \,
\ell_2\cdot P(p_4)\cdot X(\ell_1, p_1)\cdot P(p_4)\cdot \ell_2 \,.
\nonumber
\end{align}
To the order we are interested in we may freely replace $p_4\rightarrow -p_1$, since the difference is of subleading order in the expansion 
in small transferred momentum.
For $n=2$, the second and third line vanish and, for $X\equiv E$, we recover the four-point matrix element of the  operator
$ \phi { E}^{(l)}_{\mu_1\mu_2}{  E}^{(l)}{}^{\mu_1\mu_2}\phi$ given in Eqs.~\eqref{M4Esq}.

Sewing this matrix element with two three-point scalar-graviton amplitudes in Eq.~\eqref{threept} using the rule \eqref{eq:sewing_rule} 
leads to
\begin{align}
\label{AmplitudeEBmn}
{\cal M}_{{\rm X}^2_{l,n}}(\bm p,\bm q) &= 8 (8\pi G)^2
i^{2(n-2)} m_1 m_2^4
\, \frac{2 (n-2)!}{n!}
\cr
&\quad \times \Big[
 {\cal M}{}_{n}^{(l)}(\Pi_1^{ X})+ 
 2(n-2){\cal M}{}_{n}^{(l)}(\Pi_2^{ X}) 
+\frac{1}{2}(n-2)(n-3){\cal M}{}_{n}^{(l)}(\Pi_3^{ X})
\Big] \,,
\\
 {\cal M}_{l,n}(\Pi_k^{X})&= 
 \int \frac{d^D \ell}{(2\pi)^D} \frac{(u_1\cdot \ell)^{2l}((u_1\cdot \ell)^2  + \frac{1}{2} q^2 )^{n-2}}
{\ell^2 ((\ell-p_2)^2-m_2^2)  (\ell-q)^2} \left(\frac{q^2}{(u_1\cdot \ell)^2  + \frac{1}{2} q^2 }\right)^{k-1}
 { {\cal M}}(\Pi_k^{ X})  \,,
\end{align}
where $k=1,2,3$. 

Both ${ {\cal M}}(\Pi_k^{\cal E}) $ and ${ {\cal M}}(\Pi_k^{\cal B}) $
have the same general structure: 
\begin{align}
{ {\cal M}}(\Pi_i^{X})  &= 
A_i^{  X}(u_1\cdot \ell)^2((u_1\cdot \ell)^2  + \textstyle{\frac{1}{2}} q^2 )
+B_i^{  X}  q^2 (u_1\cdot \ell)^2
\cr
& \qquad\;
+C_i^{  X}  q^2((u_1\cdot \ell)^2  + \textstyle{\frac{1}{2}} q^2)
+D_i^{  X} q^4(1 - 2 \sigma^2)^2 \,.
\end{align}
The coefficients $A,\dots, D$ for the amplitude with an insertion of an electric-type operator are given by
\begin{align}
&A_1^{\cal E} = 1\,,
\quad
&&B_1^{\cal E} = -2 \sigma^2\,,
&C_1^{\cal E} = 0\,,
\quad
&&D_1^{\cal E} = \frac{1}{8}\,,~ &
\cr
&A_2^{\cal E} = \frac{1}{2} \,,
\quad
&&B_2^{\cal E} = \frac{1}{8}
(1 - 8 \sigma^2) \,,
&C_2^{\cal E} = 0\,,
\quad
&&D_2^{\cal E} = \frac{1}{16} \,, &
\cr
&A_3^{\cal E} = \frac{1}{2} \,,
\quad
&&B_3^{\cal E} = -\frac{1}{2}
\sigma^2\,,
&C_3^{\cal E} = 0
\quad
&&D_3^{\cal E} = \frac{1}{32}  &\,,
\label{Ecoef}
\end{align}
while those for the amplitude with an insertion of the ``magnetic'' operator are
\begin{align}
&A_1^{\cal B} = 4\,,
\quad
&&B_1^{\cal B} = 
(1 - 8 \sigma^2)\,,
&C_1^{\cal B} = -1\,,
\quad
&&D_1^{\cal B} = \frac{1}{2}\,, &
\cr
&A_2^{\cal B} = 2\,,
\quad
&&B_2^{\cal B} = -4 \sigma^2\,,
&C_2^{\cal B} = -\frac{1}{2} \,,
\quad
&&D_2^{\cal B} = \frac{1}{4}\,, &
\cr
&A_3^{\cal B} = 0
\quad
&&B_3^{\cal B} = \frac{1}{4}
(1 - 8 \sigma^2)\,,
&C_3^{\cal B} = -\frac{1}{4} \,,
\quad
&&D_3^{\cal B} = \frac{1}{8} & \,.
\label{Bcoef}
\end{align}

In the soft limit, all integrals in the amplitude \eqref{AmplitudeEBmn} are of the type 
\begin{align}
I_{n,2l} =\int \frac{d^D \ell }{(2\pi)^D} 
 \frac{|\bm q|^{1-2(n+l)} (u_1\cdot \ell)^{2l}((u_1\cdot \ell)^2  + \frac{1}{2} q^2 )^n}
{\ell^2 (-2u_2\cdot \ell)  (\ell-q)^2} \,;
\end{align}
they can be evaluated in terms of the triangle integrals \eqref{triangles0} found in Sec.~\ref{1loop_eg_p_space}:
\begin{align}
I_{n,2l} & = \sum_{u=1}^n C_n^u \left(-\frac{1}{2}\right)^{n-u}   I_{2(l+u)}
\cr
 & 
 = -\frac{i}{32} \frac{(-)^{n+l}}{2^{2l+n}} \frac{\Gamma(l+\textstyle{\frac{1}{2}})}{\sqrt{\pi}\Gamma(l+1)} \; 
(\sigma^2-1)^m 
{}_2F_1\left(\textstyle{\frac{1}{2}}+l, -n, 1+l, \textstyle{\frac{1}{2}}(1-\sigma^2)\right) \,,
\end{align}
where $C_n^u$ are binomial coefficients. 
In terms of these integrals, the three terms ${\cal M}{}_{n}^{(l)}(\Pi_k^{ X}) $ making up the complete amplitude are
\begin{align}
{\cal M}_{l,n}(\Pi_k^{ X}) &= 
A_k^{ X} I_{n+1 - k,2(l+1)}  +  B_k^{ X} I_{n - k,2(l+1)} +  q^2 C_k^{ X} I_{n+1-k,2l} 
+(1 - 2 \sigma^2) D_k^{ X} I_{n-k,2l})  \,,
\end{align}
with coefficients $A,\dots,D$ given in \eqref{Ecoef} and \eqref{Bcoef}. Using these building blocks it is then straightforward to 
assemble  the amplitudes ${\cal M}_{{\rm E}^2_{l,n}}(\bm p, \bm q)$ 
and ${\cal M}_{{\rm B}^2_{l,n}}(\bm p, \bm q)$ 
in Eq.~\eqref{AmplitudeEBmn}. 
The eikonal phases follows by Fourier-transforming them to impact parameter space and including the appropriate factors as in Eq.~\eqref{eikphasen=2m}. Choosing $n=2$ we recover the amplitudes in Eqs.~\eqref{AmplitudeEneq2m} and \eqref{AmplitudeBneq2m}.
Last, the two-body potential and the eikonal phase are related to the leading-order amplitude in the usual way as in Eqs.~\eqref{VvsM} and \eqref{PhasevsM}.

The position-space analysis also works in this case. In fact. for this approach it is convenient to sidestep the encoding of the tidal effects in a particular basis of higher-dimensions operators and work directly with the susceptibility $\chi$. From this perspective the matrix element of an arbitrary tidal operator quadratic in the electric field is 
\begin{align}
{\cal M}_{\chi \rm {EE}}(h(\ell_1), h(\ell_2), \phi(p_1), \phi(p_4) )
&= 2m_1\kappa^2
\chi_{\mu_1\nu_1\mu_2\nu_2}(u_1\cdot\ell_1, \hat\ell_1;u_1\cdot\ell_2, \hat\ell_2)
 {\cal E}_{\mu_1\nu_1}(\ell_1, p_1)
 {\cal E}^{\mu_2\nu_2}(\ell_2, p_1) \nn \\
&\quad + (p_1\leftrightarrow p_4, u_1\leftrightarrow u_4).
\label{eq:chielement}
\end{align}
Bose symmetry guarantees that this is symmetric in the two gravitons, so the manipulations in the previous section can be repeated here.
The Fourier transform of the one-loop integrand, after sewing the unitarity cut and evaluating the energy integral, is 
\begin{align}
\label{generalsusceptibility}
  \mathcal{M}_{\chi \rm {EE}}(\bm p,\bm r) 
&= \frac{\kappa^2}{4m_2}
\int \frac{d^{D-1}\bm \ell_1}{(2\pi)^{D-1}} \frac{e^{-i\bm r\cdot \bm \ell_1}}{\bm \ell_1^2}\,
\int \frac{d^{D-1}\bm \ell_2}{(2\pi)^{D-1}} \frac{e^{-i\bm r\cdot \bm \ell_2}}{\bm \ell_2^2}\,
{\cal M}_{\chi\rm{EE}}(h(\ell_1), h(\ell_2), \phi(p_1), \phi(p_4) )\big\vert_{\varepsilon_{\mu\nu}(\ell) \rightarrow  
	T_{\mu\nu}(p_2)} \\
&= 
\frac{m_1\kappa^4}{2m_2}
\left[\chi_{\mu_1\nu_1\mu_2\nu_2}(v \hat{\bm z}\cdot i\bm \nabla_1, \bm\nabla^{\perp}_1;v \hat{\bm z}\cdot i\bm \nabla_2, \bm\nabla^{\perp}_2)
{\cal E}_{\mu_1\nu_1}(\bm r_1, p_1) {\cal E}_{\mu_2\nu_2}(\bm r_2, p_1) \right]_{\bm r_1=\bm r_2=\bm r}
\,,
\nonumber
\end{align}
where $v=\sqrt{\sigma^2-1}$, $\bm\nabla^{\perp} = \bm\nabla - v^2 \, \hat{\bm z}\, (\hat{\bm z}\cdot \bm \nabla)$, and we have introduced different positions, $\bm r_i$, for all the gravitons. They are to be set equal after the derivatives are evaluated.
As before, we can obtain the isotropic potential by first generating the on-shell amplitude through Eq.~\eqref{eq:fourierToMom} and Fourier transforming back to the position space. 
The eikonal phase can either be obtained from Eq.~\eqref{PhasevsM} or directly from $  \mathcal{M}_{\chi \rm {EE}}(\bm p,\bm r)$ via Eq.~\eqref{eq:phaseFromPosition}.

\subsection{Adding spin}

It is not difficult to formally the calculation in the previous sections to include spin degrees of freedom for the particle with momentum $p_2$. 
It amounts to changing $T_{\mu\nu}(p_2)$ in Eqs.~\eqref{Esqsewing}, \eqref{positionspace_1loop}, \eqref{AmplitudeEBmn} 
and \eqref{eq:chielement} with $T_{\mu\nu}^\text{Kerr}(p_2, l_i)$ in Eq.~\eqref{Kerr_sewing} or its general form defined from 
Eq.~\eqref{Lstresstensor} and parametrized as in Eq.~\eqref{sewingwithspin} and multiplying the resulting amplitude 
by the product of spin-$S$ polarization tensors.

With this replacement, the contraction of two electric-type tensors ${\cal E}_{\mu_1\mu_2}(\ell_i, p_1)$ is
 \begin{align}
 \label{fullSpinSewing}
& {\cal E}_{\mu_1\mu_2}(\ell_1, p_1) {\cal E}^{\mu_1\mu_2}(\ell_2, p_1)\big\vert_{\varepsilon_{\mu\nu}(\ell_i) \rightarrow  
	T^\text{gen}_{\mu\nu}(p_2)} \\
& \quad\;
 =\frac{1}{8} 
 m_2^4 A(\ell_1) A(\ell_2)  (8 (\ell\cdot u_1)^4 + 4 (\ell\cdot u_1)^2   q^2 (1 - 4 \sigma^2) +  q^4 (1 - 2 \sigma^2)^2)\cr
& \quad\;
{-\frac{i}{4} 
m_2^3 A(\ell_1) B(\ell_2)  q^2 \sigma (-4 (\ell\cdot u_1)^2 +   q^2 (-1 + 2 \sigma^2))\, S_2[u_1, q]} \cr
& \quad\;
{+\frac{i}{2} 
m_2^3 (A(\ell_2) B(\ell_1) + A(\ell_1) B(\ell_2)) \ell\cdot u_1  \sigma (4 (\ell\cdot u_1)^2 +   q^2 
(1 - 2 \sigma^2))\, S_2[\ell, q]} \cr
& \quad\;
{-\frac{i}{4}  
m_2^3  (A(\ell_2) B(\ell_1) - A(\ell_1) B(\ell_2)) q^2 \sigma (4 (\ell\cdot u_1)^2 +  q^2 (1 - 2 \sigma^2)) \, S_2[\ell, u_1]} \cr
& \quad\;
{+\frac{1}{2} 
m_2^2 B(\ell_1) B(\ell_2) (\ell\cdot u_1)^2 (-2 (\ell\cdot u_1)^2 +   q^2 \sigma^2)\, S_2[e^\mu, q] S_2[e_\mu, \ell] 
}\cr
& \quad\;
{+\frac{1}{2} 
m_2^2 B(\ell_1) B(\ell_2) (\ell\cdot u_1)^2 (2 (\ell\cdot u_1)^2 -   q^2 \sigma^2)\, S_2[e^\mu, \ell] S_2[e_\mu, \ell] 
}\cr
& \quad\;
{+
m_2^2 B(\ell_1) B(\ell_2) \ell\cdot u_1 ((\ell\cdot u_1)^2 -  q^2 \sigma^2)\, S_2[\ell, q] S_2[u_1, q]}\cr
& \quad\;
{-\frac{1}{2} 
m_2^2 B(\ell_1) B(\ell_2) q^2 ((\ell\cdot u_1)^2 -   q^2 \sigma^2)\, S_2[\ell, p_1] S_2[u_1, q]} \cr
& \quad\;
{-
m_2^2 B(\ell_1) B(\ell_2) (\ell\cdot u_1)^2  \sigma^2\, S_2[\ell, q]^2} \cr
& \quad\;
{+\frac{1}{2} 
m_2^2 B(\ell_1) B(\ell_2) q^2 (-(\ell\cdot u_1)^2 + q^2 \sigma^2)\, S_2[\ell, u_1]^2}+ {\cal O}(q^5) \,,
\nonumber
 \end{align} 
where $\ell_1=\ell$,  $\ell_2 = q-\ell$ and 
\begin{align}
S_2[a, b] \equiv S(p_2)^{\mu\nu}a_\mu b_\nu
\,, \qquad
S_2[e^\mu, a]S_2[e_\mu, b] \equiv \eta_{\mu\nu} S(p_2)^{\mu\rho}a_\rho  S(p_2){}^{\nu\sigma} b_\sigma  \,.
\end{align}
For vanishing spin, $A(\ell_i)=1$ and  $B(\ell_i)=0$, only the first line of Eq.~\eqref{fullSpinSewing} survives and we recover Eq.~\eqref{Esqsewing}. One may expand Eq.~\eqref{fullSpinSewing} to arbitrary order in spin. For example, to first nontrivial order, 
which corresponds to inclusion of the spin-orbit interaction for particle 2, we find
 \begin{align}
 \label{eg1S}
  &{\cal E}_{\mu_1\mu_2}(\ell_1, p_1) {\cal E}^{\mu_1\mu_2}(\ell_2, p_1)\big\vert_{\varepsilon_{\mu\nu}(\ell_i) \rightarrow  
	T^\text{gen}_{\mu\nu}(p_2)} = {\cal E}_{\mu_1\mu_2}(\ell_1, p_1) {\cal E}^{\mu_1\mu_2}(\ell_2, p_1)\big\vert_{\varepsilon_{\mu\nu}(\ell_i) \rightarrow  
	T_{\mu\nu}(p_2)}    \\
& \quad\; 
+\frac{i}{4} C_{BS^1}
m_2^3 \, \sigma  (4 \ell\cdot u_1 S_2[\ell, q] + q^2\, S_2[u_1, q] ) 
(4 (\ell\cdot u_1)^2 + q^2 (1 - 2 \sigma^2))+ {\cal O}((q\cdot S)^2) \,,
\nonumber
\end{align}
where the first term on the right-hand side is given by Eq.~\eqref{EsqEval}.

It is straightforward, albeit tedious, to write out explicitly an integral representation of the amplitude by plugging in Eq.~\eqref{fullSpinSewing}
in Eq.~\eqref{Esqsewing}. We will refrain however from doing so, and rather only comment on its structure. In addition to the integrals in Eq.~\eqref{triangles0}, the spin dependence introduces also tensor integrals: 
\begin{align}
\label{tensorint}
I_l^{\mu_1\dots \mu_s} = \int\frac{d^D \ell}{(2\pi)^D} \frac{|\bm{q}|^{-2l-s+1}\ell^{\mu_1}\dots \ell^{\mu_s} 
(\ell\cdot u_1)^{l}}{ \ell^2 ( -2 \ell\cdot u_2) (\ell - q)^2 }  \,;
\end{align}
they may be parametrized as a scalar integral $I_l[w, s]$ by contracting the free indices with an arbitrary vector $w$, from which the 
desired tensor integral is extracted by taking $s$ derivatives. Note that, unlike the triangle integrals in Eq.~\eqref{triangles0}, here the 
exponent $l$ is not constrained to be even. 
To leading order in spin only the vector integral is relevant. To this order, Eq.~\eqref{eg1S} becomes:
\begin{align} 
&{\cal M}_{{\rm E}^2_{l,2}, S(p_2)}(\bm p, \bm q)  = \varepsilon_2\cdot\varepsilon_3 
{\cal M}_{{\rm E}^2_{l,2}}(\bm p, \bm q)   
\\
& \qquad\qquad
+128 (-1)^l C_{BS^1}G^2 \pi^2 \sigma |\bm q|^{2l+3} 
\Bigl( S_2[u_1, q] \big( (-1 + 2 \sigma^2) I_{2 l} + 4 I_{2 + 2 l} \big) 
 \cr
& \qquad\qquad\qquad\qquad\; 
+ 4 S_2[e_\mu, q] \big( (1 - 2 \sigma^2) I_{1 + 2 l}^\mu
- 4 I_{3 + 2 l}^\mu
\big) \Bigr) m_1 m^3_2  \varepsilon_2\cdot\varepsilon_3+ {\cal O}((q\cdot S)^2) \,.
\nonumber
\end{align}

It is not difficult to evaluate in the usual way the vector integrals, by writing them as a linear combination of $u_1, u_2$ and $q$ and solving for the coefficients in terms of the scalar triangle integrals in Eq.~\eqref{triangles0}. Alternatively, one may re-evaluate the integrals in Eq.~\eqref{triangles0}
by treating $u_1$, $u_2$ and $q$ as uncorrelated vectors, differentiate $s$ times with respect to $u_1$ and then impose $u_i^2 = 1, u_i\cdot q = 0$.
For the vector integrals we find  
\begin{align}
I^\mu_{2l+1} =  -\frac{u_1^\mu - u_2^\mu y}{y^2-1} I_{2l+2} \,.
 \end{align}
Thus, the amplitude with the first spin-dependent term for particle 2 is
 \begin{align}
 \label{spinamplitude_LO}
 &{\cal M}_{{\rm E}^2_{l,2}, S(p_2)}(\phi(p_1),\phi(p_2),\phi(p_3),\phi(p_4)) = \varepsilon_2\cdot\varepsilon_3
 {\cal M}_{{\rm E}^2_{l,2}}(\phi(p_1),\phi(p_2),\phi(p_3),\phi(p_4))  
   \\
\quad\; 
 &- C_{BS^1} G^2 \pi^{3/2} \frac{\Gamma(\textstyle{\frac{1}{2}}+l)}{ 2^{2l-5}\Gamma(3+l)} m_2^3 
 \sigma (-1 + \sigma^2)^l (-3 + (7 + 2 l) \sigma^2) |\bm q|^{3 + 2 l} S_2[p_1, (i q)]  \varepsilon_2\cdot\varepsilon_3 + {\cal O}((q\cdot S)^2) \,.
 \nonumber
 \end{align}
To extract the two-body potential in terms of the rest-frame spin it is necessary to expand the product of polarization tensors
to leading order in spin, as discussed in Ref.~\cite{Bern:2020buy}. Using the relations 
 \begin{align}
 \varepsilon_2 \cdot  \varepsilon_3  & = \left(1  - i  
\frac{\epsilon_{rs k }p_2^r p_3^s S^k}{m_2(m_2 +E({\bm p}_2))}
+ {\cal O}(\textrm{S}^2\bm q^2 )\right) + {\cal O}(q) \,, \nn\\[5pt]
 \qquad
 \epsilon^{\mu\nu\rho\sigma}
{p_1}_\mu {p_2}_\nu q_\rho {S_i}_\sigma & = (E_1+E_2)\, (\bm p\times\bm q)\cdot \bm S_i \,,
\end{align} 
the amplitude becomes 
\begin{align}
&{\cal M}_{{\rm E}^2_{l,2}, S(p_2)}(\phi(p_1),\phi(p_2),\phi(p_3),\phi(p_4)) 
\\
&= {\overline{\cal M}}_{{\rm E}^2_{l,2}}(\bm p)|\bm q|^{2l+3} 
 + \left( \frac{{\overline{\cal M}}_{{\rm E}^2_{l,2}}(\bm p) }{m_2(E_2 + m_2)} 
 +  (E_1+E_2){\overline {\cal M}}_{{\rm E}^2_{l,2}, 1}(\bm p) \right)\;|\bm q|^{2l+3} \;  i (\bm p\times \bm q) \cdot \bm S_2 
 +{\cal O}((q\cdot S)^2)\,,
 \nonumber
\end{align}
where ${\cal M}_{{\rm E}^2_{l,2}, 1} $ is the coefficient of $S_2[p_1, (i q)]$ in Eq.~\eqref{spinamplitude_LO} and, as before,  the 
bar indicates that all $\bm q$ dependence has been extracted.
The two-body potential and the eikonal phase are then extracted by three-dimensional and two-dimensional Fourier-transforms, 
in terms of their spinless counterparts and the coefficient of the spin-dependent structure in the amplitude:
\begin{align}
V_{{\rm E}^2_{l,2}, S_2}(\bm p, \bm r) & =  V_{{\rm E}^2_{l,2}}(\bm p, \bm r)
-
\frac{(\bm p\times \bm r)\cdot {\bm S}_2}{4 E_1 E_2 |\bm r |^{2l+8}}   \,  \frac{2^{4+2 l}\Gamma(4+l)}{\pi^{3/2} \Gamma(-\textstyle{\frac{3}{2}} - l)} 
  \nn \\
 & \null  \hskip 2.4 cm \times    
 \left( \frac{{\overline{\cal M}}_{{\rm E}^2_{l,2}}(\bm p) }{m_2(E_2 + m_2)} 
  +  (E_1+E_2){\overline {\cal M}}_{{\rm E}^2_{l,2}, 1}(\bm p) \right) +{\cal O}((\bm r\bm S)^2) \,,
 \label{potentialWspin}    \\
 \delta_{{\rm E}^2_{l,2}, S_2}(\bm p, {\bm b})  & =   \delta_{{\rm E}^2_{l,2}}(\bm p, {\bm b})
 +
\frac{1}{4m_1 m_2\sqrt{\sigma^2-1}}\,   \frac{(\bm p\times \bm b)\cdot {\bm S}_2}{ |\bm b |^{2l+7}}   \,  
\frac{2^{4+2 l}\Gamma(\textstyle{\frac{7}{2}} + l)}{\pi \Gamma(-\textstyle{\frac{3}{2}} - l)}  \nn\\
 & \null  \hskip 2.4 cm \times    
 \left( \frac{{\overline{\cal M}}_{{\rm E}^2_{l,2}}(\bm p) }{m_2(E_2 + m_2)} 
  +  (E_1+E_2){\overline {\cal M}}_{{\rm E}^2_{l,2}, 1}(\bm p) \right) +{\cal O}((\bm r\bm S)^2) \,.
\label{phaseWspin}  
\end{align}
 
The position-space analysis extended to include spin degrees of freedom is equally straightforward. It amounts to
substituting  in Eqs.~\eqref{massagedE} and \eqref{generalsusceptibility} the stress tensor $T_{\mu\nu}(p_2)$ by the general 
spin-dependent one in Eq.~\eqref{sewingwithspin} or, for the scattering off a Kerr black hole, with $T^\text{Kerr}_{\mu\nu}(p_2)$ in Eq.~\eqref{Kerr_sewing}. As already emphasized, $T^\text{gen}_{\mu\nu}(\ell_i, p_2)$ depends on the graviton momentum $\ell_i$ which now
makes a leading-order contribution because of the spin dependence. Nevertheless, the contribution of $T^\text{gen}_{\mu\nu}(\ell_i, p_2)$ 
can be organized as a differential operator acting on the position-space three-dimensional scalar propagator:
\begin{align}
\label{massagedEKerr}
{\cal E}_{\mu_1\mu_2}(\bm r, p_1) 
&={\cal E}_{\mu_1\mu_2}(i{\bm \nabla}, p_1)\big\vert_{\varepsilon_{\mu\nu}(\ell_i) \rightarrow  
	T^\text{gen}_{\mu\nu}(i\bm \nabla , p_2)}  \int \frac{d^{D-1}\bm \ell_i}{(2\pi)^{D-1}} \frac{e^{-i\bm r\cdot \bm \ell_i}}{\bm \ell_i^2}	\,.
\end{align}
The structure of the stress tensor \eqref{Kerr_sewing} implies that, for scattering off a Kerr black hole, the complete spin dependence 
is governed by the non-Abelian Fourier transform
\begin{align}
 \int \frac{d^{D-1}\bm \ell_i}{(2\pi)^{D-1}} \frac{e^{-i({\hat {\bm r}}  - {\hat {\bm a}})\cdot  \bm \ell_i}}{\bm \ell_i^2} \,,
 \label{eq:naft}
\end{align}
where ${\hat {\bm r}} = {\bm r}\,\id_4$ and 
${\hat {\bm a}}$ is a vector of matrices, $({\hat {\bm a}}_\sigma){}^\mu{}_\nu = \epsilon^\mu{}_{\nu\rho\sigma}a^\rho$, 
with $a$ defined in Eq.~\eqref{adef}. One may evaluate it by formally expanding the integrand in ${\hat{\bm a}}$.

On general grounds, as discussed in Ref.~\cite{Bern:2020buy}, the impulse and spin kick is computed from the eikonal phase \eqref{phaseWspin}
through the relations \eqref{spinningeikonal} agree with those computed from Hamilton's equations of motion based on the two-body potential \eqref{potentialWspin}. The same holds for the magnetic analog of Eqs.~\eqref{phaseWspin} and \eqref{potentialWspin}.

\section{Nonlinear tidal effects}
\label{LORnSection}

\begin{figure}[tb]
	\begin{center}
		\includegraphics[scale=.7]{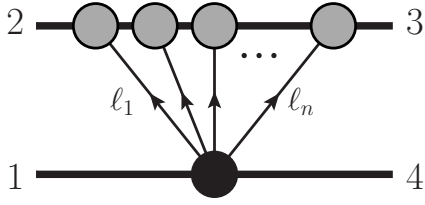}
	\end{center}
	\vskip -.3 cm
	\caption{\small The generalized cut for leading order
          contributions to nonlinear tidal operators.  Each blob
          is simply a (local) on-shell amplitude. The dark
                blob contains the $X^n$ tidal operator. The direction of graviton
                 momentum flow is indicated by the arrows. }
	\label{fig:fancut}
\end{figure}

The amplitude with nonlinear tidal effect, i.e.~the scattering with an
$X^n$ operator insertion, where $X$ stands for $E$ or $B$, can be constructed from the unitarity cut in
Fig.~\ref{fig:fancut}.  We will mostly focus on leading contribution
for such an operator in this section.  In this case, the simplifications
described in Section~\ref{LOandSpin} are all applicable.  Namely, the
amplitude with $X^n$ tidal operator is still comprised of linearized
electric and magnetic Weyl tensor in Eqs.~\eqref{calE}
and~\eqref{calB}; and the sewing of three-point amplitudes with the
amplitude with $X^n$ tidal operator is effectively replacing the
polarization $\varepsilon_{\mu\nu}(\ell_i) \rightarrow
T_{\mu\nu}(p_2)$ for each graviton.  Start from the the unitarity cut
in Fig.~\ref{fig:fancut}. After sewing we find
\begin{align}
\mathcal{M}_{{\rm X}^n}(\bm p,\bm q) =& \frac{\kappa^{n}}{m_2^{n-1}} \int
\mathcal{M}_{{\rm X}^n}(h(\ell_1),\dots,h(\ell_n),\phi(p_1),\phi(p_4))\big\vert_{\varepsilon_{\mu\nu}(\ell_i) \rightarrow T_{\mu\nu}(p_2)} \nn \\
&\quad \times
\frac{1}{\ell_1^2 \ell_2^2\cdots \ell_n^2}
\left[\frac{i}{(-2u_2\cdot 
	\ell_1)}\frac{i}{(-2u_2\cdot \ell_{12})}\dotsc 
\frac{i}{(-2u_2\cdot \sum_{j=1}^{n-1}\ell_{j})}
\right],
\label{GeneralRnAmpl}
\end{align}
where we integrate over $\ell_{i}$ with $i=1,\dots,n-1$ and $\sum^n_{i=1} l_i = q$.

As discussed in the previous section, we can include spin degrees of
freedom for the field without the tidal deformation by simply
replacing in Eq~\eqref{GeneralRnAmpl} the point-particle stress tensor
$T_{\mu\nu}$ with that of the general spinning particle
$T^\text{gen}_{\mu\nu}$, cf.~Eq.~\eqref{sewingwithspin}, or with that
of a Kerr black hole, cf.~\eqref{Kerr_sewing}.
 
The calculations from position space and momentum space also follow similarly as before. We discuss them in turn.

\subsection{Leading order position-space analysis}
\label{LORnRealSection}
Start with Eq.~\eqref{GeneralRnAmpl}. 
Again we consider the rest frame of particle 2 in which we have Eq.~\eqref{eq:restframe2}.
The first step is to integrate out energy in potential region. Using the identity~\cite{Saotome:2012vy}
\begin{align}
\delta \left(\sum_{i=1}^n \ell^0_i \right)
\left[\frac{i}{(-2u_2\cdot 
	\ell_1+i0)}\frac{i}{(-2u_2\cdot \ell_{12}+i0)}\dotsc 
\frac{i}{(-2u_2\cdot \sum_{j=1}^{n-1}\ell_{j}+i0)}
+\textrm{perm}
\right]
= \pi^{n-1} \prod_{i=1}^n \delta (\ell^0_i)\,,
\end{align}
where perm is the rest of $n!$ permutations of $\ell_{1,\dots,n}$. Since the integrand is invariant under permutations, this localizes all $\ell^0_i=0$ with a $1/n!$ prefactor
\begin{align}
\mathcal{M}_{{\rm X}^n}(\bm p,\bm q) =& \frac{(-\kappa)^{n}}{(2m_2)^{n-1}\,n!} \int
\, \left[\prod_{i=1}^{n} \frac{d^{D-1}\bm \ell_i}{(2\pi)^{D-1}}\, \frac{1}{\bm \ell_i^2} \right] \,
\delta \left(\bm q -\sum_{i=1}^n \bm \ell_i \right)
\nn \\
&\quad \times 
\mathcal{M}_{{\rm X}^n}(h(\ell_1),\dots,h(\ell_n),\phi(p_1),\phi(p_4))\big\vert_{\varepsilon_{\mu\nu}(\ell_i) \rightarrow T_{\mu\nu}(p_2)},
\label{GeneralRnAmpl_3D}
\end{align}

To evaluate this integral, we use the same manipulations as at one loop. First consider the Fourier transform to position space
\begin{align}
&\mathcal{M}_{{\rm X}^n}(\bm p,\bm r) = \int \frac{d^{D-1}\bm q}{(2\pi)^{D-1}} e^{-i\bm r \cdot \bm q} \widetilde{\mathcal{M}}_{{\rm X}^n} (\bm p,\bm q) \nn \\
&\hspace{2cm}= 
\frac{(-\kappa)^{n}}{(2m_2)^{n-1}\,n!}\prod_{i=1}^n \int \frac{d^{D-1}\bm \ell_i}{(2\pi)^{D-1}}
\frac{e^{-i\bm r\cdot \bm \ell_i}}{\bm \ell_i^2}
\,
\mathcal{M}_{{\rm X}^n}(h(\ell_1),\dots,h(\ell_n),\phi(p_1),\phi(p_4))\big\vert_{\varepsilon_{\mu\nu}(\ell_i) \rightarrow T_{\mu\nu}(p_2)} \nn \\ 
&\hspace{2cm}= \frac{(-\kappa)^{n}}{(2m_2)^{n-1}\,n!}\, 
\mathcal{M}_{{\rm X}^n}(h_1,\dots,h_n,\phi(p_1),\phi(p_4)|\bm r) ,
\label{eq:position_General}
\end{align}
where we use Eq.~\eqref{massagedE} to define
\begin{align}
\label{eq:position_comptonGeneral}
&\, \mathcal{M}_{{\rm X}^n}(h_1,\dots,h_n,\phi(p_1),\phi(p_4)|\bm r)
\\ &\hspace{2cm}\equiv
\prod_{i=1}^n \int \frac{d^{D-1}\bm \ell_i}{(2\pi)^{D-1}}
\frac{e^{-i\bm r\cdot \bm \ell_i}}{\bm \ell_i^2}
\,
\mathcal{M}_{{\rm X}^n}(h(\ell_1),\dots,h(\ell_n),\phi(p_1),\phi(p_4))\big\vert_{\varepsilon_{\mu\nu}(\ell_i) \rightarrow T_{\mu\nu}(p_2)}
  \nn \\
&\hspace{2cm}= \,
\mathcal{M}_{{\rm X}^n}(h(\ell_1),\dots,h(\ell_n),\phi(p_1),\phi(p_4)|
	{ X}_{\mu_1\mu_2}(\ell_j, p_1) \rightarrow {X}_{\mu_1\mu_2}(\bm r_j, p_1),
	\bm \ell_j \rightarrow i\nabla_j
	)\Big\vert_{\bm r_j \rightarrow \bm r}\,.\nn
\end{align}
As before all the coordinates $\bm r_j$ are identified as $\bm r$ in the end.
The above formula is very general and applies to higher multipole operators or general susceptibilities similar to Eq.~\eqref{generalsusceptibility}.
Recall that $\mathcal{M}_{{\rm X}^n}(h(\ell_1),\dots,h(\ell_n),\phi(p_1),\phi(p_4))$ is only a
function of ${\cal E}_{\mu_1\mu_2}(\ell_i, p_2)$, ${\cal
B}_{\mu_1\mu_2}(\ell_i, p_2)$, and Mandelstam invariants. 
The Fourier transform simply replaces them with their corresponding in position-space expressions defined 
in Eqs.~\eqref{massagedE} and~\eqref{massagedB}.  
As before, the result of $\mathcal{M}_{{\rm X}^n}(\bm p,\bm r)$ is generally not isotropic, because any $u_1\cdot \ell$ in momentum space
generates dependence on $\hat{\bm z} \cdot \bm \ell$. To bring it into the isotropic form, we Fourier transform back to momentum space, as in
Eq.~\eqref{eq:fourierToMom}.

A simple example is the operator ${E}_{\mu}{}^{\nu} {E}_{\nu}{}^{\rho} {E}_{\rho}\,^{\mu}$, denoted as $(E^3)$. 
With the contraction of three ${\cal E}$ tensors \eqref{massagedE} given by
\begin{equation}
{\cal E}_{\mu}{}^{\nu}(\bm r, p_1){\cal E}_{\nu}{}^{\rho}(\bm r, p_1) {\cal E}_{\rho}\,^{\mu}(\bm r, p_1) 
=\frac{3m_2^6}{4096\pi^3 |\bm r|^{13}}\, \left[9(\sigma^2-1)(\bm r^2- z^2)(\sigma^2 \bm r^2 -(\sigma^2-1)z^2)+2\bm r^2
\right] \, ,
\label{E3positionspaceXX}
\end{equation}
the graviton-scalar amplitude is
\begin{align}
\mathcal{M}_{({\rm E}^3)}(h_1,h_2,h_3,\phi(p_1),\phi(p_4)|\bm r) = 12\kappa^3\,m_1\, 
{\cal E}_{\mu}{}^{\nu}(\bm r, p_1){\cal E}_{\nu}{}^{\rho}(\bm r, p_1) {\cal E}_{\rho}\,^{\mu}(\bm r, p_1) \,.
\end{align}
Plugging into Eq.~\eqref{eq:position_General} then yields the four-scalar amplitude in position space
\begin{align}
\mathcal{M}_{({\rm E}^3)}(\bm p,\bm r) = -\frac{\kappa^6m_1}{2 m_2^2}\, 
{\cal E}_{\mu}{}^{\nu}(\bm r, p_1){\cal E}_{\nu}{}^{\rho}(\bm r, p_1) {\cal E}_{\rho}{}^{\mu}(\bm r, p_1).
\end{align}
Using the Fourier transform formula in Eq.~\eqref{eq:FT_generalTensor}, we arrive the final result
\begin{align}
{\mathcal M}_{({\rm E}^3)}(\bm p, \bm q) &  = 
\frac{-|\bm q|^{6-4\epsilon}}{2\epsilon}\,
\overline{{\mathcal M}}_{({\rm E}^3)}(\bm p) =
\frac{18}{11!!} G^3 m_1 m_2^4 \pi {} \Bigl( \frac{7}{4} - 9 \sigma^2 + 10 \sigma^4 \Bigr) \frac{|\bm q|^{6-4\epsilon}}{\epsilon} \,.
\end{align}

An important feature of the position-space scalar-graviton amplitude \eqref{eq:position_comptonGeneral}, which 
we already encountered in the one-loop analysis in Sec.~\ref{real_oneloop}, is that it factorizes into a product of 
position-space ${\cal E}$ tensor, defined in Eq.~\eqref{massagedE} and its magnetic counterpart, perhaps 
with additional derivatives. 
As explained in \sect{OperatorsSection}, the fact that these position-space tensors have rank 3 implies that such 
a product can be further expressed as a sum of products of traces of at most three factors. 
For example, Eq.~\eqref{EnvsE2E3} gives the decomposition of any power of a rank-3 matrix in terms of in 
terms of traces of two and three such matrices. It applies directly to the four-scalar amplitude with an insertion 
of $(E^{n})$ and expresses it as a sum of four-scalar amplitudes with an insertion of $(E^2)^{n_2} (E^3)^{n_3}$ 
with $n = 2 n_2 + 3 n_3$. It also applies directly to amplitudes with an insertion of $(B^{n})$. While the resulting 
amplitude vanishes of $n$ is odd, it also further simplifies if $n$ is even. The parity-odd nature of 
${\cal B}_{\mu, \nu}(\bm r, \bm p)$ and position-space factorization imply that, to leading order, $(B^3)$ = 0
because there are insufficient vectors to saturate the Levi-Civita tensor. 
Therefore, to leading order, the analog of Eq.~\eqref{EnvsE2E3} for the magnetic operators reduces to
\begin{align}
(B^{n=2k}) = \frac{1}{2^{k-1}}(B^2)^k \, .
\end{align}
The amplitudes collected in the Appendix~\ref{Appendix:summary} verify these formulas for up to $n = 8$.

The momentum-space four-scalar amplitude is related to the position-space four-scalar amplitude by single 
$(D-1)$-dimensional  Fourier transform. The structure of the position-space amplitude is essential. This 
observation allows us to evaluate amplitudes and the corresponding two-body potentials to leading order
for arbitrary operators. 

Since the position-space scalar-graviton amplitudes with one insertion of either one of $(E^2)$, $(B^2)$ or $(E^3)$
have a similar structure, we will discuss them simultaneously, referring to these operators as $(\mathcal O)$.
They have the form, 
\begin{equation}
{\widetilde {\mathcal M}}_{(\mathcal O)} = {\mathcal N}_{(\mathcal O)}
   \frac{1}{ \bm r^h}  \Bigl( a_{(\mathcal O)}  + b_{(\mathcal O)}  \frac{ (\bm r \cdot  \bm u_1)^2}{\bm r^2} 
   + c_{(\mathcal O)}  \frac{(\bm r \cdot \bm u_1)^4}{\bm r^4} \Bigr) \,,
\label{PositionXm}
\end{equation}
where $\mathcal N_{(\mathcal O)}$ is an operator-dependent normalization factor. For the three operators it is,
\begin{align}
{\mathcal N}_{(\rm E^2)}={\mathcal N}_{(\rm B^2)}=2^4 G^2 \pi^2  m_1 m_2^3 \,,
\hskip 2. cm 
{\mathcal N}_{(\rm E^3)}=2^{5} G^{3} \pi^{3} m_1 m_2^{4} \, ,
\end{align}
and the coefficients are 
\begin{align}
a_{\rm (E^2)} &=   \frac{ 3 (1 - 3 \sigma^2 + 3 \sigma^4)} {2 \pi^2} \,, \hskip 1.8 cm 
b_{\rm (E^2)} =   \frac{9 (1 - 2 \sigma^2)} {2 \pi^2} \,,   \hskip 2.1 cm       
c_{\rm (E^2)} =   \frac{9}{2 \pi^2}  \,, \nn \\
a_{\rm (B^2)} &=   \frac{ 9 \sigma^2 (\sigma^2 -1)} {2 \pi^2} \,, \hskip 2.7 cm 
b_{\rm (B^2)} =  \frac{9 (1 - 2 \sigma^2)} {2 \pi^2} \,,   \hskip 2.1 cm       
c_{\rm (B^2)} =  \frac{9}{2 \pi^2}  \,,  \nn \\
a_{\rm (E^3)} &= - \frac{3 (2 - 9 \sigma^2 + 9 \sigma^4)}{8 \pi^3} \,, \hskip 1.5 cm
b_{\rm (E^3)} = - \frac{27 (1 - 2 \sigma^2)}{8 \pi^3}\,, \hskip 1.5 cm
c_{\rm (E^3)} = - \frac{27}{8 \pi^3} \, .
\label{abcCoeffs}
\end{align}
The exponent of the overall $\bm r$ factor is $h = 6$ for $({\cal O}) = (E^2)$ and $({\cal O}) = (B^2)$ and
$h = 9$ for $({\cal O}) = (E^3)$.

The position- space amplitude with an insertion of an operator made up of $n$ such traces is simply 
given by raising \eqref{PositionXm} to the $n$th power and adjusting the normalization factor,
\begin{equation}
\label{Opowern}
{\widetilde {\mathcal M}}_{(\mathcal O)^n} = {\mathcal N}_{(\mathcal O)^n}
\Bigl[\frac{1}{\bm  r^h}  
\biggl( a_{(\mathcal O)}  + b_{(\mathcal O)}  \frac{ (\bm r \cdot \bm u_1)^2}{\bm r^2} + c_{(\mathcal O)}  \frac{(\bm r \cdot \bm u_1)^4}{\bm r^4} \Bigr) \biggr]^n \, .
\end{equation}
The change in normalization factor is related to the normalization of the tree-level amplitude with one 
insertion of the composite operator. We find
\begin{align}
{\mathcal N}_{(\rm E^2)^n}  = {\mathcal N}_{(\rm B^2)^n}  =  2^{2n+2} G^{2n} \pi^{2n} m_1 m_2^{2n+1} \, ,
\qquad
{\mathcal N}_{ (\rm E^3)^n}  =  2^{3n+2} G^{3n} \pi^{3n} m_1 m_2^{3n+1}  \,.
\end{align}

To obtained the momentum-space scattering amplitude with an insertion of an arbitrary operator $({\cal O})^n$ 
we first use twice the binomial expansion and put the position-space amplitude in the form
\begin{equation}
\label{positionspaceexpanded}
{\widetilde {\mathcal M}}_{ (\mathcal O)^n} = \frac{{\mathcal N}_{ (\mathcal O)^n}}{\bm r^{nh}}
\sum_{k=0}^n \sum_{l = 0}^k {{n}\choose{k}} {{k}\choose{l}}  
      a_{\mathcal O}^{n-k} \, b_{\mathcal O}^l \, c_{\mathcal O}^{k-l} \, 
\biggl(\frac{(\bm r \cdot \bm u_1)^2}{\bm r^2} \biggr)^{\! 2k - l} \,.
\end{equation}
Using then the general tensor Fourier-transform relation \eqref{eq:FT_generalTensor} which enforces 
$\bm q\cdot \bm u_1 = \bm q^2/2 \rightarrow 0$ leads to the desired result:
\begin{align}
{\cal M}_{ (\mathcal O)^n}(\bm p, \bm q) &= \frac{{\mathcal N}_{ (\mathcal O)^n}}{ |\bm q|^{D - n h-1}} 
\sum_{k=0}^n \sum_{l = 0}^k {{n}\choose{k}} {{k}\choose{l}}  
      a_{\mathcal O}^{n-k} \, b_{\mathcal O}^l \, c_{\mathcal O}^{k-l} 
      \\
& \qquad \times      
      \frac{ 2^{D-h n -1}  \pi^{D/2}(\sigma^2-1)^{2k-l} \Gamma(\frac{1}{2} + 2 k - l)}
      {\sin(\frac{\pi}{2}(D  - h n - 1))\Gamma(\frac{1}{2}(3 + h n - D))\Gamma(2 k - l + \frac{1}{2} h n)} \,,
\nonumber      
\end{align}
where $D=4-2\epsilon$.
The two-body potential and the eikonal 
phase follow then straightforwardly via Eqs.~\eqref{eq:Mbardef}-\eqref{eq:FTransform}:
\begin{align}
\hskip -.3 cm 
\label{potentialspaceexpanded}
V_{ (\mathcal O)^n}(\bm p, \bm r) &= -\frac{{\mathcal N}_{ (\mathcal O)^n}}{4E_1E_2\; |\bm r|^{n h}} 
\sum_{k=0}^n \sum_{l = 0}^k {{n}\choose{k}} {{k}\choose{l}}  
      a_{\mathcal O}^{n-k} \, b_{\mathcal O}^l \, c_{\mathcal O}^{k-l} \, (\sigma^2-1)^{2k-l} 
      \frac{ \Gamma(\frac{1}{2} + 2 k - l) \Gamma(\frac{1}{2} h n)}{\sqrt{\pi} \Gamma(2 k - l + \frac{1}{2} h n)}  \,,
      \\
\hskip -.3 cm 
\delta_{ (\mathcal O)^n}(\bm p, \bm b) &= \frac{{\mathcal N}_{ (\mathcal O)^n}}{4m_1m_2\; |\bm b|^{n h-1}} 
\sum_{k=0}^n \sum_{l = 0}^k {{n}\choose{k}} {{k}\choose{l}}  
      a_{\mathcal O}^{n-k} \, b_{\mathcal O}^l \, c_{\mathcal O}^{k-l} \, (\sigma^2-1)^{2k-l-1/2} 
      \frac{ \Gamma(\frac{1}{2} + 2 k - l) \Gamma(\frac{1}{2} (h n-1))}{\Gamma(2 k - l + \frac{1}{2} h n)}   \,.
      \nonumber
\end{align}

As discussed earlier, parity and factorization of the position-space amplitude implies that, to leading order
in the classical limit, amplitudes with an insertion of an operator which has at least one parity-odd factor 
vanish identically even if the operator is overall parity-even.  Thus, Eq.~\eqref{EnvsE2E3} with $E\rightarrow B$
implies that  the approach described here yields the two-body potential for all nonlinear tidal operators of the 
type $(B^{2n})$. 

The discussion above can be easily extended to cover amplitudes with
one insertion of $(E^n)$. Eq.~\eqref{EnvsE2E3} expresses it as a
linear combination of amplitudes with one insertion of
$(E^2)^{n_2}(E^3)^{n_3}$ with $2n_2+3n_3=n$.  The position space form
of the latter involves a product of two factors analogous to the
right-hand side of Eq.~\eqref{Opowern}. Each of them can be binomially
expanded (with a slight simplification based on the equality
$b_{(E^2)}/b_{(E^3)} = c_{(E^2)}/c_{(E^3)}$ visible in
\eqn{abcCoeffs}) and put in a form analogous to the right-hand side of
Eq.~\eqref{positionspaceexpanded}.  Fourier-transforming using
Eq.~\eqref{eq:FT_generalTensor} and putting together all terms leads
to the momentum-space amplitude with one insertion of $(E^n)$.

The general formulas above show explicitly that the difference $E^{2n} - B^{2n}$ is subleading in the high-energy limit. This extends the observations of Refs.~\cite{PMTidal, PortoTidalTwoLoop} beyond the linear order.

\subsection{Order by order momentum-space analysis}

The above position-space evaluation is a very effective means
for evaluating leading contributions to any given tidal operator.
Momentum-space methods for evaluating the loop integrals instead offer
a straightforward way to systematically extend the results to 
higher orders following the methods presented in
Refs.~\cite{CliffIraMikhailClassical,3PMPRL,3PMLong}.  Indeed
following these methods, next to leading order contributions to $E^2$ and $B^2$
tidal operators were evaluated in
Ref.~\cite{CheungSolonTidal}. A related approach
for tidal operators based on world lines has been recently given in
Ref.~\cite{PortoTidalTwoLoop} where additional $E^2$
operators were evaluated.

Here we first re-evaluate the amplitudes in momentum space through $C^4$ and then 
discuss the extension to higher orders. The starting point is again the generalized cut shown in Fig.~\ref{fig:fancut}.   We evaluate the expressions in
$D$-dimensions.  Here we do not make use of the special real-space factorization of the integrals discussed in the previous section, but rather simply carry out the evaluation of the cut and then reduce the result to a basis of independent momentum products. 
We can simplify the resulting expressions considerably by applying the cut conditions and expanding in small momentum transfer $q$. Specifically, we can choose a basis of momentum invariants which does not contain any of the products ($p_2\cdot \ell_k$), since the cut conditions give
\begin{align}
  \left(-p_2+\sum_{i=1}^k\ell_i\right)^2 - m_2^2= 0 \,&\rightarrow\,\,(p_2\cdot \ell_k)=\sum_{i=2}^k\sum_{j=1}^{i-1}(\ell_i\cdot \ell_j)-\sum_{i=1}^{k-1}(p_2\cdot \ell_i)\,,
\end{align}
where the final term can be eliminated inductively starting with $p_2\cdot\ell_1=0$. Products of the form ($p_3\cdot \ell_k$) can then be eliminated using momentum conservation $p_3= -p_2 -q = -p_2 - \sum \ell_i$. Since the cut graviton momenta scale as $\cO(q)$, the cut conditions thus ensure that the scaling of ($p_2\cdot \ell_k$) or ($p_3\cdot \ell_k$), which naively would be $\cO(q)$, instead scale as $\cO(q^2)$. This greatly aids in the simplification of the integrand after expanding in small $q$.

Unlike in the position-space analysis,
the integrals do not decouple into a product, and
in general, the momentum-space integrals can be challenging to evaluate. To do so, we use FIRE6~\cite{Fire} which uses integration 
by parts methods~\cite{IBP} to reduce
the integrals a single master integral, which can then be evaluated
either by direct integration or by differential equations~\cite{DEs}.
Evaluating the integrals is the most significant
bottleneck for this method, but the task is significantly aided by the
use of special variables as described in \cite{Parra-Martinez:2020dzs},
\begin{equation}
p_1=-(\bar{p}_1-q/2) \,, \hskip .8 cm  p_4 = \bar{p}_1 + q/2\,,
  \hskip .8 cm  p_2 =-(\bar{p}_2+q/2) \,, \hskip .8 cm  p_3 = \bar{p}_2 - q/2 \,.
\end{equation}
The $\bar p_i$ are orthogonal to $q$ by construction: $\bar{p}_i\cdot q = 0$.
As described in more detail in Ref.~\cite{Parra-Martinez:2020dzs}, with these variables the matter propagators reduce to
\begin{equation}
\frac{1}{\left(p_2+\ell_{1\cdots i}\right)^2-m_2^2}
= \frac{1}{2\bar{p}_2\cdot \ell_{1\cdots i}} + \cO(q^0) \,,
\label{eq:linearize}
\end{equation}
so the matter propagators are linear in the loop
momenta.  In addition, we can define normalized external momenta,
$\bar u_i^\mu = {\bar{p}_i^\mu} /{\sqrt{m_i^2 - {q^2}/ 4}}$, such that
$\bar u_i^2 = 1$ The net effect is that the $q^2$ dependence is scaled out
of the integral so that it is only a function of a single-scale
$\bar u_1 \cdot \bar u_2 = \sigma + \mathcal{O}(q^2)$.  Using these variables
integral encountered at any order of perturbation
theory can then be converted to a single scale integral.  Such
integrals are quite amenable to integration-by-parts methods, greatly
speeding the evaluation.

\begin{figure}[tb]
	\begin{center}
		\includegraphics[scale=.7]{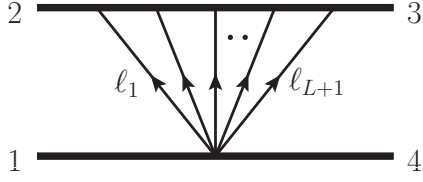}
	\end{center}
	\vskip -.3 cm
	\caption{\small The $L$-loop fan integral.}
	\label{fig:fan_int}
\end{figure}

The restriction to the potential region precludes pinching any propagators and the existence of irreducible scalar products.
Thus, the result of IBP reduction is a single master integral, with a coefficient given by powers of $\bm q$ dictated by dimensional analysis, as well as a polynomial in $\sigma$. The master integral is the scalar \emph{fan} integral in Fig.~\ref{fig:fan_int}, which can be easily evaluated by factorizing the loops by going to position space and Fourier transforming back, with the result
\begin{align}
  I_{\text{fan}}^{(L)} &= \int \left(\prod_{i=1}^{L+1}\frac{d^D\ell}{(2\pi)^D}\frac{1}{\ell_i^2}\right)\frac{|\bm q|^{2-L}\delta(\sum_i\ell_i -q)}{(-2u_2\cdot \ell_1 + i0)(-2u_2\cdot \ell_{12} + i0)\cdots(-2u_2\cdot \ell_{1\cdots n-1} + i0)}\nonumber \\
  &=  \frac{i^{L+2}  }{2^{ L (4-2\epsilon) } \pi^{L\left(\frac{3}{2}-\epsilon\right)}} \frac{\Gamma \left(\frac{1}{2}-\epsilon\right)^{L+1} \Gamma
   \left((\epsilon-\frac{1}{2}) L+1\right)}{\Gamma(L+2) \Gamma \left( (\frac{1}{2} - \epsilon)
   (L+1)\right)} |\bm q|^{-2\epsilon L} \,. 
\end{align}
At one loop this agrees with Eq.~\eqref{eq:oneloopint} with $l=0$, and at two and three loops it yields 
\begin{equation}
 I_{\text{fan}}^{(2)} = 
 \frac{1}{768 \pi ^2  } \frac{(\bm q^2)^{-2 \epsilon }}{2\epsilon} + \mathcal{O}(\epsilon^0)\,, \hskip 1.5 cm 
 I_{\text{fan}}^{(3)} = -\frac{i}{49152 \pi ^2} + \mathcal{O}(\epsilon)\,.
\end{equation}

The results of the IBP reduction at two loops gives the amplitudes
with a single insertion of the tidal operators in terms of a single
master integral:
\begin{align}
\cM_{\rm (E^3)}&= 
 \frac{1024}{385 } \pi^3 G^3 m_1 m_2^4 |\bm q|^6 
   \left(\frac{7}{4}-9 \sigma ^2 + 10 \sigma ^4\right) I_{\text{fan}}^{(2)} 
\,,\nn\\
\cM_{\rm (EB^2)}&=
 \frac{1024}{1155 } \pi^3  G^3  m_1 m_2^4 |\bm q|^6 \left(\sigma^2-1\right)
       \left(1+10\sigma^2  \right) I_{\text{fan}}^{(2)} 
  \,,\nn\\
  \cM_{\rm (B^3)}&= \cM_{\rm E^2B} = 0 \,.
\end{align}
As expected, the parity
odd operators $E^2B$ and $B^3$ operator do not contribute.

At three loops, by reducing the integrand to the sole master integral
we find the following for the amplitudes with an insertion of the
single trace operators, 
\begin{align}
\mathcal M_{\rm (E^4)} & =
- i \frac{ 983} {9031680}  \pi^4 G^4 m_1 m_2^5 |\bm q|^{9} 
(1231 - 7304 \sigma^2 + 18590 \sigma^4 - 22880 \sigma^6 + 12155 \sigma^8)  I^{(3)}_{\text{fan}}
 \,, \nn \\
\mathcal M_{\rm (B^4)} & = 
- i \frac{140569} {9031680} \pi^4  G^4  m_1 m_2^5 |\bm q|^{9}
(\sigma^2 -1 )^2 (1 + 10 \sigma^2 + 85 \sigma^4) I^{(3)}_{\text{fan}}
\,, \nn \\
\mathcal M_{\rm (EEBB)} & = 
-i \frac{10813} {27095040} \pi^4 G^4 m_1  m_2^5
  |\bm q|^{9} (\sigma^2-1) (41 + 689 \sigma^2 - 2925 \sigma^4 + 3315 \sigma^6) I^{(3)}_{\text{fan}}
 \,, \nn\\
\mathcal M_{\rm (EBEB)} & = 
i \frac{10813}{27095040} \pi^4 G^4 m_1 m_2^5 |\bm q|^{9} (\sigma^2-1) 
(25 + 481 \sigma^2 - 2925 \sigma^4 + 3315 \sigma^6) I^{(3)}_{\text{fan}}
 \,.
\end{align}
Similarly, the amplitudes with double trace insertions evaluate to,
\begin{align}
\mathcal M_{\rm (E^2)^2} &  = - i \frac{ 983} {4515840}  \pi^4 G^4 m_1 m_2^5 |\bm q|^{9} 
(1231 - 7304 \sigma^2 + 18590 \sigma^4 - 22880 \sigma^6 + 12155 \sigma^8)  I^{(3)}_{\text{fan}} \,, \nn \\
\mathcal M_{\rm (B^2)^2} & =- i \frac{140569} {4515840} \pi^4  G^4  m_1 m_2^5 |\bm q|^{9}
(\sigma^2 -1 )^2 (1 + 10 \sigma^2 + 85 \sigma^4) I^{(3)}_{\text{fan}} \,, \nn \\
\mathcal M_{\rm (E^2) (B^2)} & =
 - i \frac{10813} {4515840} \pi^4 G^4  m_1 m_2^5 |\bm q|^{9}
    (\sigma^2 -1) (19 + 299 \sigma^2 - 975 \sigma^4 + 1105 \sigma^6)  I^{(3)}_{\text{fan}}
 \,,\nn\\
\mathcal M_{\rm (E B)^2} & = 0 \,,
\end{align}
It is not difficult to check that these results satisfy the four-dimensional
relations described in \sect{OperatorsSection}. In addition, they agree
with the results obtained in the previous section for tidal operators
with arbitrary numbers of $E$s and $B$s and collected in the Appendix for
a variety of operators up to $E^8$ and $B^8$.

\begin{figure}[tb]
	\begin{center}
		\includegraphics[scale=.7]{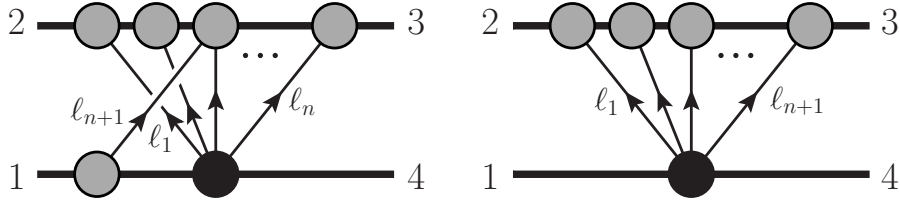}
	\end{center}
	\vskip -.3 cm
	\caption{\small The generalized cuts that need to be evaluated at next
          to leading order for an $R^n$ type tidal operator.  
	}
	\label{fig:NLOcuts}
\end{figure}

\begin{figure}[tb]
	\begin{center}
		\includegraphics[scale=.7]{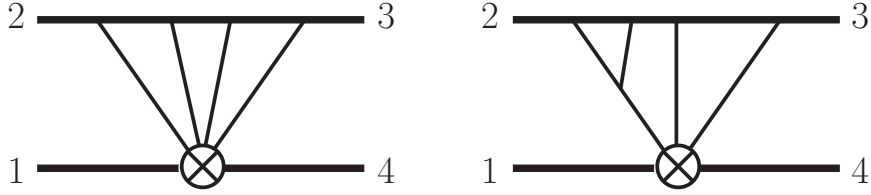}
	\end{center}
	\vskip -.3 cm
	\caption{\small Sample diagrams for next-to-leading-order contributions for the 
             $R^3$ tidal operators which are simple to evaluate. }
	\label{fig:NLODiags1}
\end{figure}

\begin{figure}[tb]
	\begin{center}
		\includegraphics[scale=.7]{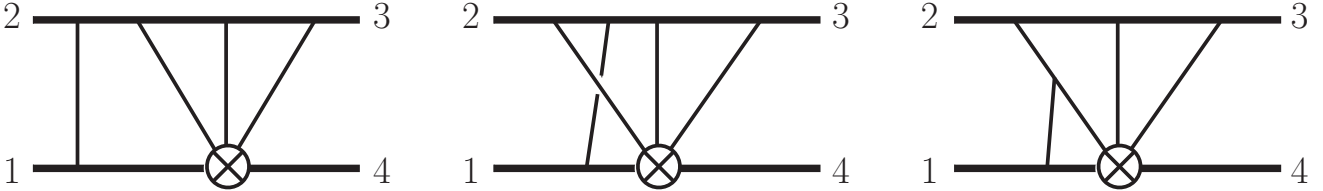}
	\end{center}
	\vskip -.3 cm
	\caption{\small Sample diagrams next-to-leading order contributions 
            for the $R^3$ tidal operators that involve iteration contributions 
            or nontrivial integrals.  
	}
	\label{fig:NLODiags2}
\end{figure}

An important aspect of the momentum-space approach is that it gives a
systematic means for obtaining corrections higher order in Newton's
constant for any operator insertion.  For example \fig{fig:NLOcuts}
shows the generalized cuts that would need to be evaluated to obtain
the next-to-leading order corrections from an $C^3$ tidal operator.
In the first of these cuts the four-point amplitude can appear at any
location on the top matter line. The mapping of the integrands
resulting from these cuts onto a integral basis generates a number of
diagrams. For example, in \fig{fig:NLODiags1} we show a sample of the
diagrams that that are quite easy to evaluate for an $R^3$ tidal
operator, as we can again evaluate the integral using the real-space
technique presented in the previous section.  More complicated
diagrams that involve iteration contributions or non-trivial
integrations are shown in \fig{fig:NLODiags2}. In these cases, the
integrals do not factorize, but the momentum-space approach of
evaluating cuts and reducing to a basis of master integrals will still
be quite feasible. As noted in Refs.~\cite{CheungSolon3PM,
  CheungSolonTidal} the probe limit simplifies the evaluation of the
contributions.  In any case, it is clear that amplitude methods can be
applied beyond leading order to understand the systematics of
higher-dimension operators.  We leave this to future studies.


\section{Effective field theory extensions of GR}
\label{LOPureRnSection}

The same methods apply just as well to any operator, not just the
tidal ones.  For example, we can consider the $R^n$ operators arising
from unknown short distance physics.  Here we will not classify such
operators, but pick illustrative examples.  The effect of operators up
to $R^4$ has already been discussed in some detail in
Refs.~\cite{pureR4, pureR3, pureR3Other}.  In order to be concrete
here we discuss an effective action of the form
\begin{equation}
S=\frac{1}{16\pi G}\int d^Dx \sqrt{-g}\left(-R+c_K K_{\mu_1...\rho_n} 
R^{\mu_1\nu_1\sigma_1\rho_1} R^{\mu_2\nu_2\sigma_2\rho_2} \cdots R^{\mu_n\nu_n\sigma_n\rho_n}\right)\,,
\end{equation}
where the first term is the usual Einstein-Hilbert action, and
$K_{\mu_1...\rho_n}$ merely gives the contraction between the Riemann
tensors. Each independent contraction carries an independent Wilson
coefficient $c_K$.

\begin{figure}
	\centering \includegraphics[scale=.75]{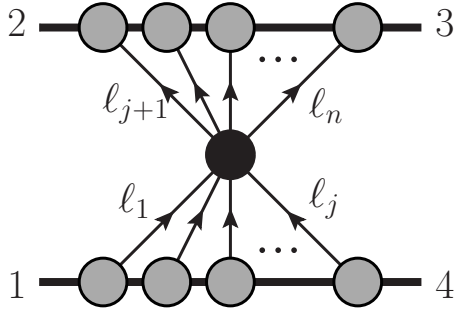} \caption{Cut
		for a general $R^n$ type operator. In the case $j=1$, it is
		convenient to take the single graviton attaching to the bottom
		matter line as off shell and part of a tree amplitude
		including the lower massive scalar line. All other gravitons
		and exposed matter lines are taken
		on shell. The direction of graviton momentum flow is indicated by the arrows.
		\label{fig:Rncut}
	}
\end{figure}	

We construct the integrands for pure $R^n$ modifications of gravity in
a similar manner as for those of the tidal operators. The leading
contribution to the potential due to $R^n$ operators is captured by
the cuts in \fig{fig:Rncut}. The diagrams in general are a product of
two fan diagrams, where all graviton legs, as well as the matter lines
between the three point vertices, are on shell, the exception being
the case where $(n-1)$ on-shell gravitons attach to one of the matter
lines, while one graviton which we take to be off shell attaches to
the other matter line. In this case, it is convenient to include the
matter line to which the single graviton propagator is attached as part
of a single tree amplitude.

To evaluate the cuts in \fig{fig:Rncut} we use the replacement derived
above (see \eqn{eq:sewing_rule}). This simplifies the form of the
Riemann tensor:
\begin{equation}
R^{\mu\nu\rho\sigma}(\ell_i) \big\vert_{\varepsilon_{\mu\nu}(\ell_i) \rightarrow  
	T_{\mu\nu}(p_a)} =
- \frac{1}{2} \left(\ell_i^\mu \ell_i^\rho \left(p_a^\nu p_a^\sigma -\frac 12 \eta^{\nu\sigma}m_a^2 \right) - (\sigma \leftrightarrow \rho)\right) + ( (\mu,\rho) \leftrightarrow (\nu,\sigma))  + \mathcal O(q^3)  \,, \label{eq:Rreplaced}
\end{equation}
where $p_a$ and $m_a$ are the momentum and mass of the matter line the
graviton attaches to. When contracted in sequence with other gravitons
attaching to the same matter line, products involving the matter
momenta in the above expression must reduce to $p_a \cdot p_a =
m_a^2$, or the $q$ scaling will become sub-leading, as shown in the
previous section.

The cut corresponding to Fig.~\ref{fig:Rncut} is simply a product of two fans, 
\begin{equation}
\cC_{R^n}
=K_{\mu_1...\rho_n}\cO^{\mu_1...\rho_j}(\ell_1, ..., \ell_j;p_1)  \cO^{\mu_{j+1}...\rho_n}(\ell_{j+1}, ..., \ell_n;p_2) \,,
\label{eq:CRn}
\end{equation}
where, for instance,
\begin{equation}
\cO^{\mu_1...\rho_j}(\ell_1, ..., \ell_j;p_1) =R_1^{\mu_1\nu_1\sigma_1\rho_1}\cdots R_j^{\mu_j\nu_j\sigma_j\rho_j}\big\vert_{\varepsilon_{\mu\nu}(\ell_i) \rightarrow  
	T_{\mu\nu}(p_1)} \,.
\end{equation}
As in previous sections, the integrands obtained after restoring the cut
propagators are also well suited for applying position-space
techniques. In this case, we must introduce a fictitious momentum
transfer $\bm q'$ such that the integrand decouples in two parts,
corresponding to the two terms in Eq.~\eqref{eq:CRn} decouple, and the
corresponding propagators attached to one matter line or the other.
The energy integrations can be carried out as in the previous sections
with the result
\begin{align}
\mathcal{M}_{R^n}(\bm p, \bm q) &= K_{\mu_1...\rho_n}\int d^{D-1} \bm q' \, \delta(\bm q+\bm q') 
\int \left(\prod_{a=1}^j \frac{d^{D-1}\bm \ell_a}{(2\pi)^{D-1}}\right)\frac{\delta(\sum_{a=1}^j \bm \ell_a + \bm q') \cO^{\mu_1...\rho_j}(\bm\ell_1, ..., \bm\ell_j;p_1)}{\bm \ell_1^2\cdots\bm \ell_j^2} \nn \\
&\hspace{2cm}\times\int \left(\prod_{a=j+1}^n \frac{d^{D-1}\bm \ell_a}{(2\pi)^{D-1}}\right)\frac{\delta(\sum_{a=j+1}^n \bm \ell_a - \bm q) \cO^{\mu_{j+1}...\rho_n}(\bm\ell_{j+1}, ..., \bm\ell_n;p_2)}{\bm \ell_{j+1}^2\cdots\bm \ell_n^2} \,.
\end{align}
Writing
\begin{equation}
\delta(\bm q+\bm q') = \int \frac{d^{D-1}\bm x}{(2\pi)^{D-1}} e^{i(\bm q + \bm q')\cdot \bm x}
\end{equation}
and taking the Fourier transform of the amplitude we find
\begin{align}
\mathcal{M}_{R^n}(\bm p, \bm r) &= \int \frac{d^{D-1}\bm q}{(2\pi)^{D-1}} e^{-i\bm q\cdot \bm r} \mathcal{M}_{R^n}(\bm p, \bm q) \nn \\
& = K_{\mu_1...\rho_n}\int d^{D-1}\bm x
\int \left(\prod_{a=1}^j \frac{d^{D-1}\bm \ell_a}{(2\pi)^{D-1}} \frac{e^{-i \bm \ell_a \cdot \bm x }}{\bm \ell_a^2}\right) \cO^{\mu_1...\rho_j}(\bm\ell_1, ..., \bm\ell_j;p_1) \nn \\
&\hspace{1cm}\times\int \left(\prod_{a=j+1}^n \frac{d^{D-1}\bm \ell_a}{(2\pi)^{D-1}} \frac{e^{-i \bm \ell_a \cdot (\bm r-\bm x) }}{\bm \ell_a^2} \right) \cO^{\mu_{j+1}...\rho_n}(\bm\ell_{j+1}, ..., \bm\ell_n;p_2) \nn \\ 
&= K_{\mu_1...\rho_n}\int d^{D-1}\bm x  \,    \cO^{\mu_1...\rho_j}(\bm x;p_1)      \cO^{\mu_{j+1}...\rho_n}(\bm r-\bm x;p_2) \,.
\label{eq:Rnxspace}
\end{align}
The product in momentum space has become a convolution in position
space over $\bm x$, which can be viewed as the position in the bulk,
i.e. away from the massive particle trajectories, at which the $R^n$
operator is inserted. Note however that this formula does not have a
natural interpretation in position space, given that the energy
integrals in each factor were performed by going to the rest frame of
different particles.  In practice, as in previous sections, this
formula can be used by transforming one last time to momentum space,
so that the convolution is trivialized and each factor can be written
in isotropic coordinates.

The inclusion of derivatives, $\nabla^{2m}R^n$, or of spin on the
matter lines poses no obstruction to applying this method.  In the
former case one must organize the additional powers of loop momentum
in the integrand into either factor in analogy with Eq.~\eqref{eq:CRn}.
The factorization argument carries over and the additional loop
momenta become derivatives in position space acting on either factor
of Eq.~\eqref{eq:Rnxspace}.
For the case of spin, the only difference is that the Fourier transforms in \eqn{eq:Rnxspace} become non-Abelian Fourier transforms defined in Eq.~\eqref{eq:naft}.

\begin{figure}[tb]
	\begin{center} \includegraphics[scale=.6]{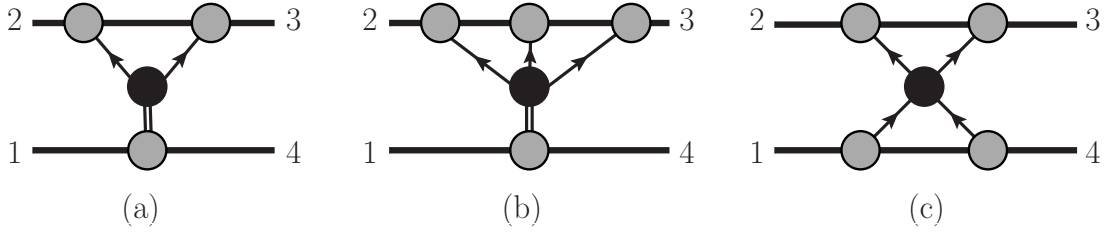} \end{center} \vskip
	-.3 cm \caption{\small The corrections from (a) $R^3$
		and (b,c) $R^4$ operators that appear in EFT extensions of GR. The double-line notation
		indicates that we have not used on-shell conditions on
		that line.   
	} \label{fig:R4cuts}
\end{figure}

As simple examples, consider the cases of $\cO_{R^3}=R^{\mu_1\nu_1}{}_{\mu_2\nu_2}R^{\mu_2\nu_2}{}_{\mu_3\nu_3}R^{\mu_3\nu_3}{}_{\mu_1\nu_1}$ and $\cO_{(R^2)^2}=(R^{\mu_1\nu_1}{}_{\mu_2\nu_2}R^{\mu_2\nu_2}{}_{\mu_1\nu_1})^2$.
The contributing generalized unitarity cut for the $R^3$ operator are
shown in \fig{fig:R4cuts}(a) while the two potentially contributing
cuts for the the $R^4$ operator are shown in \fig{fig:R4cuts}(b,c). In
the diagrams the double-line notation indicates that we have not used
on-shell conditions on that line, but consider the two connected blobs
as part of a single tree amplitude.\footnote{Whether on-shell
	conditions are used on the intermediate leg corresponds to shifting
	the coefficient of $\phi R^n \phi$ operators.}

After carrying out the integration, the $R^3$ and $R^4$  amplitudes are 
\begin{align}
\cM_{R^3}=&
- 6 c_{R^3}  G^{2} \pi^2  m_1^2m_2^2(m_1+m_2)|\bm q|^3( \sigma^2 - 1 ) 
\,, \nn \\
%
\cM_{(R^2)^2}=&
-\frac{2^7}{315} c_{(R^2)^2}  G^{3} \pi
m_1^2 m_2^2 (m_1^2+m_2^2)\frac{({\bm q}^2)^{3 -2\eps} }{2 \eps}( 3 \sigma^2 - 1)   
\,,
%
\end{align}
where we took the operators to have coefficient $c_R^3$ and $c_(R^2)^2$ respectively.  Taking the Fourier transform \eqref{eq:FTransform} to position space 
gives the potentials
\begin{align}
V_{R^3} & = \frac{18}{E_1 E_2} c_{R^3}  G^{2} 
m_1^2m_2^2(m_1+m_2) ( \sigma^2 - 1) \, \frac{1}{r^6} \,, \nn \\
V_{(R^2)^2} & = \frac{2^8}{ E_1 E_2} c_{(R^2)^2} G^{3} \,
m_1^2 m_2^2 (m_1^2+m_2^2) ( 3 \sigma^2 - 1 ) \, \frac{1}{r^9} \,.
\end{align}

The
$\cO_{R^3}$ amplitude and potential was obtained previously in Refs.~\cite{pureR3,
	pureR3Other} and we find agreement. In Ref.~\cite{pureR3} the authors also evaluate the
effect of an additional $R^3$ operator,
\begin{equation}
G_3=\cO_{R^3}-R^{\mu\nu\alpha}{}_{\beta}R^{\beta\gamma}{}_{\nu\sigma}R^{\sigma}{}_{\mu\gamma\alpha}   \,;
\end{equation}
this is related to tidal operators via a field redefinition up to
operators that vanish in four dimensions.  This can be seen by
evaluating its four-dimensional four-point amplitude, which feeds into
the two-graviton cut, using spinor-helicity methods~\cite{pureR3}:
\begin{equation}
\cM_{G^3}(\phi(p_1),h^{++}(k_2),h^{++}(k_3),\phi(p_4)) \propto [23]^4(-q^2+2m_1^2) \,.
\end{equation}
Since this is a local contribution, it is already captured by tidal
operators of the form $E^2$, $B^2$. Interestingly, though, if this
operator were present with a sufficiently large coefficient, it would
produce a result equivalent to the leading tidal Love numbers, even if
these are set identically to zero for black holes in Einstein
gravity~\cite{LoveZero}.

The leading PN contribution from the $R^4$ operator ($\cO_{(R^2)^2}$) was
calculated in Ref.~\cite{pureR4}, with which we find agreement. We can also easily determine that the other operators considered in Ref.~\cite{pureR4} give no contribution to the leading conservative potential. The contribution from $\cO_{(R^2)(R\tilde R)}=(R^{\mu_1\nu_1}{}_{\alpha\beta}\epsilon^{\alpha\beta}{}_{\mu_2\nu_2}R^{\mu_2\nu_2}{}_{\mu_1\nu_1})(R^{\mu_3\nu_3}{}_{\mu_4\nu_4}R^{\mu_4\nu_4}{}_{\mu_3\nu_3})$ is zero simply because it is parity-odd. The operator $\cO_{(R\tilde R)^2}=(R^{\mu_1\nu_1}{}_{\alpha\beta}\epsilon^{\alpha\beta}{}_{\mu_2\nu_2}R^{\mu_2\nu_2}{}_{\mu_1\nu_1})^2$, while being parity even, contributes zero at leading order, in analogy to the tidal operator $\cO_{(EB)^2}$. In both cases, the factorization of the integrand in real space forces the separate parity-odd factors to evaluate to zero, as discussed in Section \ref{LORnRealSection}

Here we refrain from evaluating the amplitudes for the $R^5$ and
higher operators. However, in these cases, there is an additional link
between the $R^n$ extensions of Einstein gravity and the tidal
operators.  After carrying out the soft expansion of the integrand for
the $R^n$ operators, one encounters ultraviolet divergences that
renormalize tidal operators~\cite{NRGR}. For example, in principle the
$R^5$ operator, which produces a diagram with three gravitons attached
to one matter line and two attached to the other, could produce a UV
subdivergence and thereby renormalize $E^2$ or $B^2$ tidal operators (with additional derivatives). It
would  be an interesting problem to systematically study this interplay
for infinite sequences of $R^n$ operators.

\section{Conclusions}
\label{ConclusionSection}

In this paper we evaluated the leading-PM order contributions to the two-body Hamiltonian 
from infinite classes of tidal operators using momentum space and position space scattering
amplitude and effective field theory methods.
The same principles yield leading-PM order Hamiltonian terms from tidal deformations
probed by a spinning particle and also from effective field theory modifications of
general relativity. 
Our results offer a new perspective on the general structure of linear and nonlinear tidal effects in the relativistic 
two-body problem while also being of potential phenomenological interest.

Our analysis of $E^2$ and $B^2$ tidal operators arbitrary number of derivatives is similar 
to that of Ref.~\cite{Haddad:2020que}, except that we use a basis of operators which aligns 
with the more standard worldline tidal operators~\cite{NRGR, PMTidal}.  Their Wilson coefficients
are the same (up to an overall normalization that we provide) with the worldline electric and 
magnetic tidal coefficients which in turn are proportional to the corresponding multipole 
Love numbers.
By directly evaluating all relevant integrals we obtain explicit expressions for the two-body 
Hamiltonian and the amplitude's eikonal phase, from which both scattering and closed-orbit 
observables can be found straightforwardly.
We illustrated the inclusion of spin by working out the leading-order tidal contributions from 
$E^2$-type operators with arbitrary number of derivatives for one object interacting with the 
spin of the other.  

For tidal operators with arbitrary numbers of electric or magnetic components of the Weyl tensor, the 
integrand for the leading-order contributions are not difficult to construct because their building 
blocks are tree-level leading order on-shell matrix elements  of the point-particle energy-momentum 
tensor and of the tidal operator.  
The simple loop-momentum dependence and the permutation symmetry of the three-point amplitude 
factors makes the integrals simple to evaluate. Indeed, Fourier-transforming all graviton propagators
decouples all integrals from each other, making it straightforward to write down explicit results for 
infinite classes of tidal operators.
We have verified that the results obtained this way thought direct momentum space integration.
While position space methods make leading-order calculations straightforward, momentum-space 
methods can be applied systematically, to arbitrary PM order.

An interesting feature of gravitational tidal operators, which we exploited in their description, is their 
close similarity with gauge theory operators describing the interaction of extended charge distributions
with electromagnetic fields. This formal connection extends to dynamical level double-copy relations. 
For leading-order contributions this is a straightforward consequence of the factorization of the 
linearized Riemann tensor into two gauge-theory  field strengths and of the factorization of the 
energy-momentum tensor into two gauge theory currents.  
Such double-copy factorizations also hold for the energy-momentum tensor~\cite{Bern:2020buy}.  It 
would be very interesting to investigate double-copy relations beyond the leading PM order.

In summary, in this paper we took some steps towards systematically evaluating contributions to the 
two-body Hamiltonian from infinite families of tidal operators.  The leading order in $G$ results
are remarkably simple, suggesting that much more progress will be forthcoming.

\subsection*{ Acknowledgments:}
We are especially grateful for discussions with Clifford Cheung, Nabha
Shah, and Mikhail Solon for discussions and sharing a draft of their
article with us.  We also thank Dimitrios Kosmopoulos, Andreas Helset and Andr\'es Luna for discussions. 
Z.B. and E.S. are supported by the U.S. Department of Energy
(DOE) under award number DE-SC0009937.  
J.P.-M. is supported by the U.S. Department of Energy (DOE) under award number~DE-SC0011632.
R.R.~is supported by the U.S. Department of Energy (DOE) under grant number~DE-SC0013699.
C.-H.S.~is grateful for support by the Mani L. Bhaumik Institute for
Theoretical Physics and by the U.S. Department of Energy (DOE) under award number~DE-SC0009919.

\newpage
\appendix

\section{Appendix: Summary of Explicit Results}
\label{Appendix:summary}

In this appendix we collect explicit results for scattering amplitudes
with a tidal operator insertion.  Using \eqn{VvsM}, this immediately
gives us the potential.  Here we consider the amplitudes with operator
insertions of the type $E^{n-2m} B^{2m}$.  We express the amplitude in
terms of the variable $\sigma = p_1 \cdot p_2/m_1 m_2$.  
The general formulae for $(E^2)^n$, $(B^2)^n$ and $(E^3)^n$
are given from Eq.~\eqref{positionspaceexpanded} to Eq.~\eqref{potentialspaceexpanded} with the coefficients in Eq.~\eqref{abcCoeffs}.
Here we give
explicit results corresponding up to 7 loops in the amplitudes
approach.  As noted in the text, the amplitudes with an odd $B$-field
insertions vanish by parity so we do not include those.  We also do
not explicitly list cases where a trace contains an odd number of $B$s
since these also vanish.

To list the amplitudes we scale out the powers of $| \bm q|$ from the scattering amplitudes,
following \eqn{eq:Mbardef}, 
\begin{equation}
{\mathcal M}_{{\rm X}^{2n}} =  |\bm q|^{3 (2n-1)} \overline{\mathcal M}_{{\rm X}^{2n}} 
=|\bm q|^{3 (2n-1)} \textrm{C}_{{\rm X}^{2n}}
\,,
\end{equation}
for a tidal operator which we build from a total of $2n$ $E$s or $B$s,
independent of the trace structure.  
For operators where total number
of $E$s and $B$ is odd the rescaling is bit difference because of the
appearance of a divergence
\begin{equation}
{\mathcal M}_{{\rm X}^{2n+1}} 
= |\bm q|^{6n - 4 n \eps } \overline{\mathcal M}_{\rm X^{2n}}
= - \frac{1}{2n} \frac{1}{\eps} |\bm q|^{6n - 4 n \eps }\, \textrm{C}_{\rm X^{2n+1}} \,,
\end{equation}
The long-range classical contribution comes from the $\log \bm q^2$ term that arises from expanding in $\eps$.

As discussed in \sect{OperatorsSection}, the potential is given in the two-body Hamiltonian is given
by a Fourier transform \eqref{VvsM} and the eikonal phase is also given by Eq.~\eqref{PhasevsM}.
Carrying out the Fourier transform we have from \eqn{eq:VMbar} and \eqn{phaseV1}
\begin{align}
V_{{\rm X}^{2n}} &= 
-\frac{1}{4E_1 E_2} \frac{8^{2n-1}\,\Gamma(3n)}{\pi^{3/2}\Gamma(\frac{3}{2}-3n)}\frac{\textrm{C}_{{\rm X}^{2n}}}{|\bm r|^{6n}}\,, \\
\delta_{{\rm X}^{2n}} &= 
\frac{1}{4m_1 m_2 \sqrt{\sigma^2-1}} \frac{8^{2n-1}\,\Gamma(3n-\frac{1}{2})}{\pi \Gamma(\frac{3}{2}-3n)}\frac{\textrm{C}_{{\rm X}^{2n}}}{|\bm b|^{6n-1}}\,,
\end{align}
where we only keep the finite term in $\eps$.
Similarly, for the odd powers 
\begin{align}
V_{{\rm X}^{2n+1}} &= 
\frac{1}{4E_1 E_2} \frac{(-1)^{n}\,\Gamma(6n+2)}{2\pi}\frac{\textrm{C}_{{\rm X}^{2n+1}}}{|\bm r|^{6n+3}}\,, \\
\delta_{{\rm X}^{2n+1}} &= 
\frac{1}{4m_1 m_2 \sqrt{\sigma^2-1}} \frac{(-1)^{n-1}8^{2n}\,\Gamma(3n+1)^2}{\pi}\frac{\textrm{C}_{{\rm X}^{2n+1}}}{|\bm b|^{6n+2}}\,.
\end{align}

\noindent
For $X^2$ we have,
\begin{align}
\textrm{C}_{\rm (E^2)} &  = 
\frac{5}{2^3} G^2 m_1 m_2^3 \pi^2 \Bigl(\frac{11}{5} - 6 \sigma^2 + 7 \sigma^4 \Bigr) 
\,, \nn \\
\textrm{C}_{\rm (B^2)} &  = 
\frac{5}{2^3} G^2 m_1 m_2^3 \pi^2 (\sigma^2-1) \Bigl(1+ 7 \sigma^2 \Bigr)
\,,
\end{align}
where the parenthesis on the operator denote the matrix trace, as defined in \eqn{Parentheses}

\noindent
For $X^3$:
\begin{align}
\textrm{C}_{\rm (E^3)} &  = 
-\frac{2^2\, 3^2}{11!!} G^3 m_1 m_2^4 \pi {} \Bigl( \frac{7}{4} - 9 \sigma^2 + 10 \sigma^4 \Bigr)
 \,, \nn \\
\textrm{C}_{\rm (EB^2)} &=
- \frac{2^2\, 3}{11!!} G^3 m_1 m_2^4 \pi {} (\sigma^2 -1) \Bigl(1 +  10 \sigma^2 \Bigr)
 \,.
\end{align}

\noindent
For $X^4$:
\begin{align}
\textrm{C}_{\rm (E^4)} &  = 
-\frac{11\cdot 13}{ 2^{12}\, (7!!)^2} G^4 m_1 m_2^5 \pi^2 
  \Bigl(\frac{1231}{143} - \frac{664}{13} \sigma^2 + 130 \sigma^4 - 160 \sigma^6 + 85 \sigma^8 \Bigr) 
\,, \nn\\
\textrm{C}_{\rm (B^4)} &  = 
-\frac{11\cdot 13}{2^{12} \, (7!!)^2} G^4 m_1 m_2^5 \pi^2 (\sigma^2-1)^2 \Bigl(1 + 10 \sigma^2 + 85 \sigma^4 \Bigr) 
\,, \nn \\
\textrm{C}_{\rm (EEBB)} & =
 -\frac{11 \cdot 13}{2^{12}\, (7!!)^2 }
 G^4 m_1 m_2^5 \pi^2   (\sigma^2 -1) \Bigl(\frac{41}{39} + \frac{53}{3} \sigma^2 - 75 \sigma^4 + 85 \sigma^6 \Bigr) 
 \,, \nn\\
\textrm{C}_{\rm (EBEB)} & = 
\frac{11 \cdot 13}{2^{12} \, (7!!)^2}
G^4 m_1 m_2^5 \pi^2  (\sigma^2 -1) \Bigl(\frac{25}{39} + \frac{37}{3} \sigma^2 - 75 \sigma^4 + 85 \sigma^6 \Bigr) 
\,, \nn \\
\textrm{C}_{\rm (E^2)^2} &  = 
2 \textrm{C}_{\rm (E^4)}
 \,, \nn \\
\textrm{C}_{\rm (B^2)^2} & = 
2 \textrm{C}_{\rm (B^4)}
 \,, \nn \\
\textrm{C}_{\rm (E^2) (B^2)} & = 
-  \frac{11 \cdot 13}{2^{11} \, (7!!)^2} G^4 m_1 m_2^5 \pi^2 
(\sigma^2 - 1) \Bigl(\frac{19}{13} + 23 \sigma^2 - 75 \sigma^4 + 85 \sigma^6 \Bigr) 
 \,.
\end{align}

\noindent
For $X^5$:
\begin{align}
\textrm{C}_{\rm (E^5)} &  = 
\frac{1}{2^6\, (19!!)} G^5 m_1 m_2^6 \pi {}
   \Bigl(1094 - 8535 \sigma^2 + 24608 \sigma^4 - 32832 \sigma^6 + 17280 \sigma^8 \Bigr)
 \,, \nn\\
\textrm{C}_{\rm (E^3 B^2)} & = 
\frac{1}{2^6\, 5 \,  (19!!)} G^5 m_1 m_2^6 \pi {}
(\sigma^2 -1) \Bigl(499 + 10144 \sigma^2 - 46656 \sigma^4 + 51840 \sigma^6 \Bigr) 
 \,, \nn\\
\textrm{C}_{\rm (EBEBE)} & = 
-\frac{1}{2^5\, 5 \,  (19!!)}G^5 m_1 m_2^6 \pi {}
(\sigma^2 -1) \Bigl(61 + 1336 \sigma^2 - 7776 \sigma^4 + 8640 \sigma^6 \Bigr)
\,, \nn\\
\textrm{C}_{\rm (EB^4)} & = 
\frac{3^2}{2^2 \, 5 \, (19!!)} G^5 m_1 m_2^6 \pi {}
(\sigma^2 -1)^2 \Bigl(1 + 12 \sigma^2 + 120 \sigma^4 \Bigr)
 \,, \nn\\
%
%
\textrm{C}_{\rm (E^3) (B^2)} &  = 
\frac{3}{2^4 \, 5 \, (19!!)} G^5 m_1 m_2^6 \pi {}
   (\sigma^2 - 1) \Bigl(61 + 1336 \sigma^2 - 7776 \sigma^4 + 8640 \sigma^6 \Bigr) 
\,, \nn \\
\textrm{C}_{\rm (E^2) (EB^2)} & = 
\frac{3}{2^5 \, 5 \, (19!!)} G^5 m_1 m_2^6 \pi {}
   (\sigma^2 - 1) \Bigl(85 + 1600 \sigma^2 - 5184 \sigma^4 + 5760 \sigma^6\Bigr)
 \,, \nn \\
\textrm{C}_{\rm (B^2) (EB^2)} & =
 2\textrm{C}_{\rm EB^4}
\,.
\end{align}

\noindent
For $X^6$, $X^7$, $X^8$: 
\begin{align}
\textrm{C}_{\rm (E^6)} &  = 
\frac{17 \cdot 19 \cdot 3^5}{2^{21}\, 5^2 \, (13!!)^2} G^6 m_1 m_2^7 \pi^2
\Bigl(\frac{5558245}{26163} - \frac{328930}{171} \sigma^2 + \frac{609305}{81} \sigma^4 - \frac{144980}{9} \sigma^6 \nn \\
& \hskip 1 cm
+ \frac{183425}{9} \sigma^8 - 14950 \sigma^{10} + 5175 \sigma^{12} \Bigr)
\,, \nn\\
\textrm{C}_{\rm (B^6)} &  = 
\frac{17 \cdot 19 \cdot 3^5}{2^{21}\, 5^2 \, (13!!)^2} G^6 m_1 m_2^7 \pi^2 (\sigma^2-1)^3
     \Bigl(5 + 69 \sigma^2 + 575 \sigma^4 + 5175 \sigma^6 \Bigr)
 \,, \nn\\
\textrm{C}_{\rm (E^7)} &  = 
-\frac{3}{2^{12} \, (31!!)} 
G^7 m_1 m_2^8 \pi {}
\Bigl(1496063 - 15991430 \sigma^2 + 71940660 \sigma^4 \nn \\
& \hskip 1 cm
- 177188000 \sigma^6 + 253373120 \sigma^8 
- 200648448 \sigma^{10} + 69189120 \sigma^{12} \Bigr)
 \,, \\
\textrm{C}_{\rm (E^8)} &  = 
-\frac{23 \cdot 29 \cdot 3^7 \cdot 5}{2^{31} \, 7^2 \, (19!!)^2}
G^8 m_1 m_2^9 \pi^2 \Bigl(
\frac{57426585223}{7293645} - \frac{10076129056}{105705} \sigma^2 + \frac{32319394660}{63423} \sigma^4 \nn \\
& \hskip 1 cm
 - \frac{1227512720}{783} \sigma^6 + \frac{82520830}{27} \sigma^8 - 3916416 \sigma^{10} + 
   3294060 \sigma^{12} \nn \\
& \hskip 1 cm
- 1718640 \sigma^{14} + 441595 \sigma^{16} \Bigr) 
 \,, \nn \\
\textrm{C}_{\rm (B^8)} &  = 
-\frac{23 \cdot 29 \cdot 3^7 \cdot 5}{2^{31} \, 7^2 \, (19!!)^2}
G^8 m_1 m_2^9 \pi^2 (\sigma^2 -1)^4 \Bigl(35 + 620 \sigma^2 + 6138 \sigma^4 + 47740 \sigma^6 + 441595 \sigma^8 \Bigr) 
\,. \nn
\end{align}
As noted in \sect{LORnSection}, in the high-energy limit, where $\sigma$
is large, simple relations are visible between amplitudes with $E^2$ and $B^2$
operators inserted~\cite{PMTidal,PortoTidalTwoLoop}.

\newpage

\end{document}